%% file: LCFT_4D5D.tex
\documentclass[aps,pre,twocolumn,showpacs,superscriptaddress,groupedaddress]{revtex4} 
\usepackage{graphicx}  
\usepackage{dcolumn}   
\usepackage{bm}        
\usepackage{amssymb}   
\usepackage{tikz}
\usepackage{float}
\usepackage{xcolor}
\usepackage{esvect}
\usepackage{amsmath}
\usepackage{diagbox}
\input{mycmd}

\usepackage[vcentermath]{youngtab}
\usepackage{ytableau}
\usepackage{algorithm}
\usepackage[noend]{algpseudocode}
\usepackage{mathtools}
\usepackage{multirow}
\usepackage{hyperref}

\newcommand{\bx}{{\bf x}}

\begin{document}

\title{$N$-cluster correlations in four- and five-dimensional percolation}
\author{Xiaojun Tan}
\affiliation{Hefei National Laboratory for Physical Sciences at Microscale and Department of Modern Physics, University of Science and Technology of China, Hefei, Anhui 230026, China}
\affiliation{CAS Center for Excellence and Synergetic Innovation Center in Quantum Information and Quantum Physics, University of Science and Technology of China, Hefei, Anhui 230026, China}

\author{Youjin Deng}
\email{yjdeng@ustc.edu.cn}
\affiliation{Hefei National Laboratory for Physical Sciences at Microscale and Department of Modern Physics, University of Science and Technology of China, Hefei, Anhui 230026, China}
\affiliation{CAS Center for Excellence and Synergetic Innovation Center in Quantum Information and Quantum Physics, University of Science and Technology of China, Hefei, Anhui 230026, China}

\author{Jesper Lykke Jacobsen}
\email{jesper.jacobsen@ens.fr}
\affiliation{Laboratoire de Physique de l'\'Ecole Normale Sup\'erieure, ENS, Universit\'e PSL, CNRS, Sorbonne Universit\'e, Universit\'e de Paris, Paris, France}
\affiliation{Sorbonne Universit\'e, \'Ecole Normale Sup\'erieure, CNRS, Laboratoire de Physique (LPENS), 75005 Paris, France} 
\affiliation{Institut de Physique Th\'eorique, Universit\'e Paris Saclay, CEA, CNRS, 91191 Gif-sur-Yvette, France}

\date{\today}
\begin{abstract}
    We study $N$-cluster correlation functions in four- and five-dimensional (4D, 5D) 
    bond percolation by extensive Monte Carlo simulation.
    We reformulate the transfer Monte Carlo algorithm for percolation [Phys. Rev. E {\bf 72}, 016126 (2005)] using the disjoint-set data structure,
    and simulate a cylindrical geometry $L^{d-1}\times \infty$, with the linear size  up to $L=512$ for 4D and $128$ for 5D. 
    We determine with a high precision all possible $N$-cluster exponents, for $N \! =\!2$ and $3$, and the 
    universal amplitude for a logarithmic correlation function.
    From the symmetric correlator with $N \! = \!2$, we obtain the correlation-length critical exponent as 
    $1/\nu  \! =\! 1.4610(12)$ for 4D and $1/\nu  \! =\! 1.737 (2)$ for 5D, 
    significantly improving over the existing results.
    Estimates for the other exponents and the universal logarithmic amplitude have not been reported before to our knowledge. 
    Our work demonstrates the validity of logarithmic conformal field theory and adds to the growing knowledge 
    for high-dimensional percolation.
\end{abstract}
\maketitle

\section{Introduction}
Percolation~\cite{BroadbentHarmmersley57} 
is a cornerstone of the theory of critical phenomena~\cite{StaufferAharony1994},
and a central topic in probability theory~\cite{Grimmett1999,BollobasRiordan2006}.
The bond percolation corresponds to the $Q \! \to \! 1$ limiting case in the context of the Fortuin-Kasteleyn cluster representation
of the $Q$-state Potts model~\cite{Potts,FK},
and provides a simple yet vivid illustration of many important concepts for the latter. 
For a seminal review, see Ref.~\cite{FYWu}.
In two dimensions (2D), 
the algebraic use of symmetries---lattice duality~\cite{KW41}, Yang-Baxter integrability~\cite{Lieb67,Baxter72} and local 
conformal invariance~\cite{BPZ84,FQS84}---lead to a host of exact results for 2D systems including percolation. 
The bulk critical exponents
 $\beta$ = 5/36 (for the order parameter) and $\nu$ = 4/3 (for the correlation length) are predicted 
by Coulomb-gas arguments~\cite{Nienhuis1987}, conformal field theory~\cite{Cardy1987} and stochastic Loewner evolution
theory~\cite{LawlerSchrammWerner2001}, and are rigorously proved in the specific case of triangular-lattice site percolation~\cite{Smirnov2001}.
Above the upper critical dimensionality $d_{\rm u}$ = 6,
 the mean-field values $\beta$ = 1 and $\nu$ = 1/2 are believed to hold~\cite{Aharony1984,HaraSlade1990,Fitzner2017}.
For dimensions $2 < d < 6$, exact values of the critical exponents are still unavailable, and  
their estimates rely on numerical methods or perturbative methods ~\cite{Gracey2015}. 
Monte Carlo (MC) simulation remains a primary numerical method~\cite{Wang13,Xu2014,Paul2001}.

Apart from $\beta$ and $\nu$, there are many other critical exponents for percolation.
At percolation threshold $p_c$, a variety of fractal dimensions are used to characterize  the power-law scaling of 
the sizes of percolation clusters, the hulls,  the external perimeters, the backbones 
and the shortest paths, etc~\cite{StaufferAharony1994, Stanley1987},
and a set of exponents is defined to account for the algebraic decay of connectivity probabilities (correlation functions) 
that two far-away regions are connected by a number of mono- or poly-chromatic paths~\cite{Aizenman1999,Smirnov2001,Vincent2011}.
Even in 2D, exact values of some of these exponents are still unknown.
Recently,  a family of $N$-cluster correlation functions has been studied 
in the framework of the logarithmic conformal field theory (LCFT)~\cite{Vasseur2012, Vasseur2014, Couvreur2017,tan2019observation}.
For integer $N\geq2$,  one considers connectivity probabilities that $N$ distinct clusters 
propagate from a small neighborhood $\mathcal{V}_i$ to another one $\mathcal{V}_j$ far away,
and for each $N$, constructs a family of correlation functions from the representation theory of the symmetric group.
At $p_c$, these correlation functions decay algebraically as functions of the distance $r$, 
governed by  a set of $N$-cluster exponents,
and the amplitudes can exhibit rich behavior under rotations.
In addition, it is predicted that a certain combination of correlation functions with $N \! = \! 2$ depends logarithmically on distance, 
instead of as the usual power law, and the universal amplitude is closely related to the logarithmic coupling 
or indecomposability parameter in LCFT~\cite{gurarie2004,mathieu2007, vasseur2011indecomposability, gurarie2002conformal}.
In 2D,  the exact values of $N$-cluster exponents have been identified for $N=2,3,4$ with the help of MC simulation,
and in 3D, high-precision numerical estimates are available for $N=2,3$~\cite{tan2019observation}.
Similar results have also been obtained for the universal logarithmic amplitude.
In 2D and 3D, the universality of the logarithmic amplitude was checked via simulations on different lattices.

In this work, we extend Ref.~\cite{tan2019observation} to four and five dimensions.
In MC study of percolation, one usually
measures correlation functions in the torus geometry
, i.e., a  
$d$-dimensional hypercube with periodic boundary conditions in each of the $d$ directions~\cite{huangwei2018}. 
One major limitation of this choice is that the required computer memory grows rapidly with system linear size as $\sim L^d$, 
making it hard to simulate large system for large $d$. 
For example, simulation of a 4D~(5D) hypercube of linear size $L$=$192$~(68) needs more than five gigabyte (GB) memory.
On the other hand, a high-precision determination of percolation thresholds and of critical exponents would request data for large system sizes. 
We alleviate this problem by adopting the transfer Monte Carlo algorithm (transfer MC algorithm) in Ref.~\cite{deng2005} to effectively simulate
an infinitely long cylinder of size $L^{d-1}\times\infty$ in $d$ dimensions, 
in which each layer corresponds to a $(d-1)$-dimensional hypercube with periodic boundary condition.

The main strategy of the transfer MC algorithm, which can be regarded as a variant of the celebrated Hoshen-Kopelman (HK)
algorithm~\cite{hkalgo},
is to iteratively add a  layer of $L^{d-1}$ lattice sites during each MC step.
Only the information about connected components, i.e., to which cluster each site belongs,
of the current and newly added layers is stored in computer memory. 
The reduction of computer memory from ${\cal O}(L^d)$ to ${\cal O}(L^{d-1})$ enables one to simulate much larger systems,  
$L({\rm 4D})=512$  with two GB memory 
and  $L({\rm 5D})=128$ with four GB memory.
Moreover, the disjoint-set data structure, a simple and well-known data structure in computer science~\cite{galler1964}, 
can be well implemented to efficiently update the connectivity information
when occupied bonds are sequentially added to the cylindrical system~\cite{tarjan1979class, tarjan1984worst}. Previously, 
the disjoint-set data structure has been widely used in simulation of percolation~\cite{newman2000, newman2001,danziger2020}.
Thus, we reformulate the transfer MC algorithm in Ref.~\cite{deng2005} using the disjoint-set data structure.

We remark that, while the transfer-matrix technique~\cite{blote_nightingale}--a powerful research tool in statistical mechanics---also studies a cylindrical geometry,  
 the transfer MC algorithm is a MC sampling method by definition.
In the former, all possible configurations are summed up 
when one goes from one layer to the next, so the results are exact for the given size $L$ once 
the probabilities have exponentially converged. 
The price is that the required computer memory grows exponentially fast as $L$ increases, 
and thus the use of the transfer-matrix method is normally restricted to two-dimensional systems.

We extensively simulate critical 4D and 5D bond percolation,
and by finite-size scaling analysis, determine with a high precision all possible $N$-cluster exponents  for $N=2$ and 3 and the universal amplitude for 
the logarithmic correlation function. 
For $N=2$, the exponent for the symmetric correlation function  reduces to the two-arm exponent, 
which for the case of percolation, is also related to the red-bond exponent $y_{\rm red}$ and the correlation-length exponent $\nu$.
In the renormalization group treatment, the latter further relates to the thermal renormalization exponent as $y_{\rm t}=1/\nu$.
We obtain the correlation-length exponent as $1/\nu  \! =\! 1.4610(12)$ for 4D and $1/\nu  \! =\! 1.737 (2)$ for 5D, 
which are consistent and 
significantly improve over the most recent results $y_{\rm t} ({\rm 4D}) = 1/\nu = 1.459(6)$ and $y_{\rm t} ({\rm 5D}) = 1.747(5)$ in Ref.~\cite{Koza16}. 
The reliability of our estimates and quoted errors are carefully examined, 
and it is suggested that the thermal exponent $y_t$ in 5D is unlikely to be the central value $1.747$ reported in Ref.~\cite{Koza16}.
In addition, our estimates agree well with the 
preliminary results
$y_{\rm t} ({\rm 4D}) =1.453(37)$ and $y_{\rm t} ({\rm 5D}) = 1.741(9)$~\cite{Borinsky2020},
which are obtained from the $\phi^3$ computations up to five-loop order. 

The remainder of this work is organized as follows.
Section~\ref{algo} describes in detail the reformulation of the transfer MC algorithm,
and Sec.~\ref{quan} defines $N$-cluster correlation functions and the logarithmic correlation functions.
The simulation details and the fitting method are explained in Sec.~\ref{simu},
and the results are presented in  Sec.~\ref{result}.
Finally, a discussion is given in Sec.~\ref{discc}.

\section{Algorithm}
\label{algo}
\begin{figure}
	\includegraphics[width=\linewidth]{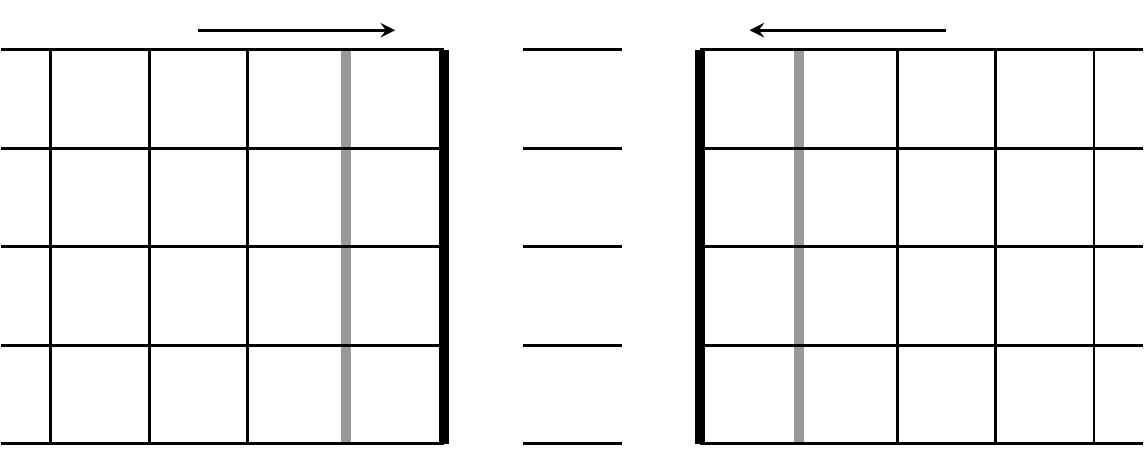}
    \caption{Sketch of the transfer MC algorithm for $d$-dimensional percolation, 
                which uses ${\cal O}(L^{d-1})$ computer memory with $L$ the side length.
                Given the connectivity of the current layer (thick gray lines),
                a layer of $L^{d-1}$ lattice sites (thick black lines) is  added 
                and the  connectivity information is determined from the current layer and  the randomly placed occupied bonds. 
                Repeating this operation effectively leads to the simulation of a half-infinitely long cylinder $L^{d-1}\times \infty$, with the newly added layer being a free surface.
                To study the bulk behavior, a pair of such cylinders is simultaneously simulated, 
                and after the two surface layers are backed up, they are ``glued" 
                into a bulk system by adding a further layer of randomly occupied bonds in-between
                (thin black lines).}
    \label{fig_1}
\end{figure}

\begin{figure*}[t]
\includegraphics[width=0.88\textwidth]{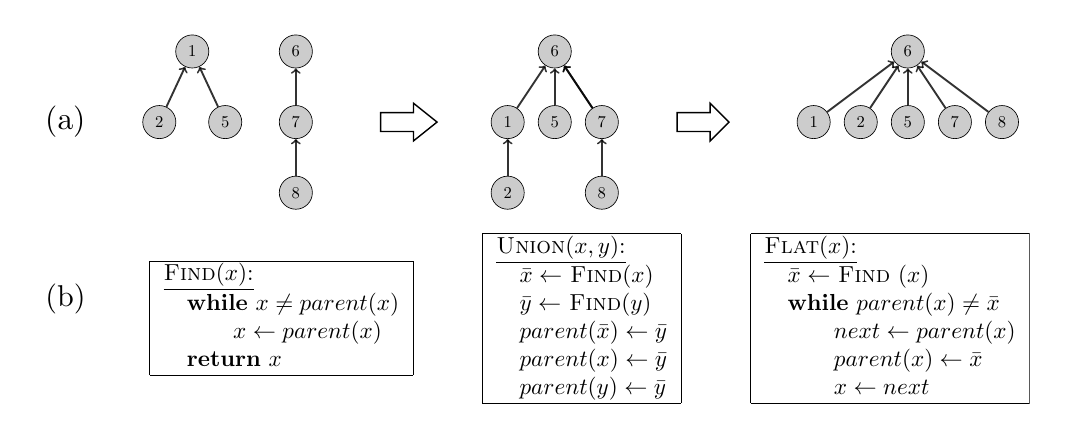}
\caption{Illustration of the tree-like disjoint-set data structure and the \textsc{Find}, \textsc{Union} and \textsc{Flat} operations. 
    In (a), there are at first two trees (clusters), one with root 1 and three sites $\{1,2,5\}$ and the other with 
    root 6 and three sites $\{6,7,8\}$. A lattice site is a root iff it points to itself; 
    for brevity, the root arrow is not shown. 
    When a bond is placed between site 5 and site 6, 
    the two clusters are merged together by operation \textsc{Union}, which finds the roots (1 and 6) by \textsc{Find} and 
    makes sites 1,5,6 to point to root 6. 
    From time to time, operation \textsc{Flat} is applied throughout the lattice to minimize the depths of trees.
    In (b), the pseudocode of operations  \textsc{Find}, \textsc{Union} and \textsc{Flat}.
}
    \label{fig_2}
\end{figure*}

In this section we shall describe in detail the transfer MC algorithm in Ref.~\cite{deng2005}
in the language of bond percolation on a $d$-dimensional hypercubic lattice of side length $L$.
Analogous procedures can be readily obtained for  site percolation.
The main strategy of the algorithm is sketched in Fig.~\ref{fig_1}.
A layer of $L^{d-1}$ lattice sites is added during each MC step,
and repeating this operation leads to an infinitely long cylinder $L^{d-1} \times \infty$.
We refer to the direction along the cylinder as the transfer direction $t$ and the perpendicular ones as the spatial directions $\bx$,
and thus each lattice site is specified by its coordinate $({\bf x}, t)$.

Each bond is occupied with probability $0\leq p \leq 1$, and two sites connected through a chain of occupied bonds are said to 
be in the same connected component, which is also called a percolation cluster. 
Occupied bonds are randomly placed with probability $p$ within the new layer $t+1$  and between the $t$th and $(t+1)$th layers.
The connectivity of the $(t+1)$th layer is solely determined by the connectivity of the $t$th layer, and the newly placed  bonds.   
Therefore, we only need to store the connectivity of the two most recent layers.  This reduces the memory cost 
from ${\cal O}(L^{d})$ to ${\cal O}(L^{d-1})$, enabling one to simulate much larger systems. 

Note that the cylinder is actually half-infinite, 
since the $t$th-layer connectivity is affected only from historical layers $t'<t$. 
In other words, the current layer $t$ is a free surface.
To study bulk behavior, we simultaneously simulate two such cylinders, and then 
``glue" the two free surfaces
into a bulk system by adding a further layer of randomly occupied bonds in-between, as illustrated in Fig.~\ref{fig_1}. 
A caution is that before the ``gluing," the connectivities of the free surfaces should be backed up  
to keep growing the cylinders.

The disjoint-set data structure, a simple tree-like data structure, supporting two simple operations---\textsc{Union} and \textsc{Find}, 
can be used to maintain the connectivity information. 
For convenience, in the actual coding we specify a lattice site by an integer $x \in [1,2V]$ 
with $x \leq V$ for the $t$th layer and $x > V$ for the $(t+1)$th layer,
where  $V=L^{d-1}$ is the volume of each layer. 
As shown in  Fig.~\ref{fig_2}, each site $x$ has a ``parent" lattice site $x'$,  indexed by $x'=parent (x)$.
All sites in the same percolation cluster form a tree graph, which is uniquely identified by the tree root.
A lattice site is a root iff it points to itself, $x=parent(x)$.
Thus, given any site $x$, the label of the percolation cluster can be easily found by following index $parent (x)$ until the root is reached,
as illustrated by operation \textsc{Find}$(x)$ in Fig.~\ref{fig_2}.

For the transfer MC algorithm, occupied bonds are added along the transfer direction $t$ sequentially.
When a bond is added between site $x$ and $y$, we apply the operation \textsc{Union}$(x,y)$ to change
the connectivity information: making one of the two roots be the ``parent" site of the other.
Meanwhile, a trick is adopted:  the new root is set to be the ``parent" sites of both $x$ and $y$,
as illustrated in Fig.~\ref{fig_2}.
The core ingredient for using the disjoint-set data structure is then the \textsc{Union}-\textsc{Find} algorithm~\cite{galler1964}.
We mention that the tree-like data structure ignores the cycle information about multi-connectivities in percolation clusters,
and thus cannot be directly applied if there are bond-deletion operations~\cite{huangwei2018}.

The computational complexity of the \textsc{Union}-\textsc{Find} method mainly depends on operation \textsc{Find}($x$)
for finding the label of a cluster, as determined by the depths of trees. 
There exist many techniques that can be utilized to avoid deep trees, 
such as path compressing and union by tree depth or tree size etc~\cite{tarjan1979class, tarjan1984worst}.
Certain combinations of these techniques are guaranteed to provide near-constant-time complexity 
on average to merge two clusters~\cite{tarjan1979class, tarjan1984worst}.
In our implementation of \textsc{Union}($x,y$), we simply make the root of $y$ be the ``parent" site of the root of $x$,
which is convenient for propagating the connectivity information from the $t$th to the $(t+1)$th layer.  
Further, at the end of constructing connectivity for an entire layer, operation \textsc{Flat}$(x)$ is taken to 
make all trees be fully flat--i.e., with minimal depth.
In the simulation, we find that it suffices to take the simple implementation in Fig.~\ref{fig_2}.

With these operations, we reformulate the transfer MC algorithm into the following steps.  

\textbf{Step 1:} Construct clusters in the new layer.
For each of the two cylinders, we take a new layer of $V=L^{d-1}$ lattice sites,  sequentially visit each pair of neighboring sites
and place an occupied bond with probability $p$, and construct percolation clusters within the layer. 
The pseudocode is shown in Step~\ref{step1}, where ``rand(~)" draws a uniform random number in $[0,1)$.

\floatname{algorithm}{Step}
\begin{algorithm}[H] 
	\caption{Construct clusters in the new layer}
\begin{algorithmic}
	\For{$x=V+1$ to  $2V$  }  \Comment{initialize, $V=L^{d-1}$}
	\State $parent(x)\leftarrow x$ 
	\EndFor
	\State
	
	\For{$x=V+1$ to $2V$ }
	\For{$k=1$ to $(d-1)$ }
	\State $y \leftarrow$ the $k$th neighbor of $x$ 
	\If {rand(~) $<p $}
		\State  \textsc{Union}($x$,$y$)
	\EndIf 
	\EndFor
	\EndFor

	\State
	
	\For{$x=V+1$ to  $2V$  } 
	\State \textsc{Flat}$(x)$ \Comment{make each tree fully flat}
	\EndFor
\end{algorithmic}\label{step1}
\end{algorithm}

\textbf{Step 2:} Add the new layer to the cylinder.
The new layer of lattice sites, where the percolation clusters are already constructed, 
is added to the cylinder by randomly placing bonds between the $t$th and $(t\!+\!1)$th layers. 
The pseudocode is given in Step~\ref{step2} and is obviously similar to Step~\ref{step1}.
A key ingredient is to incorporate the connectivity information of the $t$th layer into that of the $(t+1)$th layer, 
which is naturally realized by our implementation of operation \textsc{Union}$(z,x)$.
This leads to an important property that every site in the $(t+1)$th layer points to some site also from the $(t+1)$th layer.
Therefore, after Step~\ref{step2}, all the sites from the $(t+1)$th layer \emph{alone} form a self-contained disjoint-set data structure,
and  the $t$th-layer connectivity becomes obsolete and can be simply discarded.
 
\floatname{algorithm}{Step}
\begin{algorithm}[H]
	\caption{Add the new layer to the cylinder}
\begin{algorithmic}
	\For{$x=V+1$ to $2V$ }  
    	\State $z\leftarrow x-V$ \Comment{$z$ is in the $t$th layer}
	\If {rand(~) $<p$}
		\State  \textsc{Union}$(z,x)$ \Comment{root of $z$ points to root of $x$}
	\EndIf 
	\EndFor
	
	\State
	\For{$x=V+1$ to  $2V$  } \Comment{flat trees in the $(t+1)$th layer }
	\State \textsc{Flat}$(x)$
	\EndFor
\end{algorithmic}\label{step2}
\end{algorithm}

\textbf{Step 3:} Print the connectivity onto the $t$th layer.
The self-contained connectivity of the newly added $(t+1)$th layer
is printed onto the $t$th layer, as in Step~\ref{step3}.
By repeating Steps~\ref{step1}, \ref{step2}, \ref{step3}, one can grow a half-infinite cylinder along the transfer direction.

\floatname{algorithm}{Step}
\begin{algorithm}[H]
	\caption{Print the connectivity on the $t$th layer}
\begin{algorithmic}
	\For{$x=V+1$ to $2V$ } 
	\State   $z\leftarrow x-V$  
        \State   $parent(z)\leftarrow parent(x)- V$
	\EndFor
\end{algorithmic}\label{step3}
\end{algorithm}

\textbf{Step 4:} Glue two free surfaces 
into a bulk system.
The layer in the front of each of the two cylinders is a free surface, and the
study of bulk behavior can be achieved by gluing the two free surfaces 
into a bulk system.
After Step~3, the ``gluing" operation can be readily applied to the two $(t+1)$th layers, without affecting the $t$th-layer connectivity. By a slight modification of Step~2, one obtains two bulk layers (the former surface layers) on which the $N$-cluster correlation functions can be sampled. 

The above-formulated transfer MC algorithm can be further optimized from several aspects. 
For instance, instead of adding layer by layer, one can grow the cylinder by adding site by site, 
analogous to the sparse-matrix factorization in the traditional transfer-matrix technique. 
This can further save the computer memory, particularly if one is only interested in the surface properties. 
For high-dimensional percolation, the occupation probability $p$ is small near criticality, 
and one can apply the trick of cumulative probability to use fewer random numbers~\cite{huangwei2018}.

\section{Sampled Quantities}
\label{quan}
We use the transfer MC algorithm of Sec.~\ref{algo} to simulate bond  percolation on 4D and 5D hyper-cubic lattices,
with periodic boundary conditions in each of the perpendicular directions. 
Measurements take place in the bulk layers, 
and all sites considered below are within the same bulk layer. 
The observables are the same as those in Ref. \cite{tan2019observation} and shall be explained for completeness. 

Let $ {\mathcal V}_i \equiv (i_1, i_2, \ldots, i_N)$ denote $N$ lattice sites in a small neighborhood.
We usually take their positions to be aligned,
$\bfr_{i_{m+1}} \! = \! \bfr_{i_m} + \boldsymbol\delta$, with $m \! = \! 1,2,\ldots,N-1$. 
For $|\boldsymbol\delta|=1$, $i_m$ and $i_{m+1}$ are 
nearest neighbors. Let another site set 
${\mathcal V}_j \equiv (j_1, j_2, \ldots, j_N)$ be distant from 
${\mathcal V}_i$ 
by $\bfr = \bfr_j-\bfr_i$, with $r=|\bfr| \gg 1$. 
We consider events in which $N$ distinct percolation 
clusters propagate from ${\mathcal V}_i $  to  ${\mathcal V}_j $,
i.e., each cluster connects a site in ${\mathcal V}_i $ 
to another site in ${\mathcal V}_j $.
There are $N!$ such events, symbolically represented  
as $\PPa$ and $\PPb$ for $N=2$, 
$\PPPa$, $\PPPb$, $\PPPc$, $\PPPd$, $\PPPe$ 
and $\PPPf$ for $N=3$, etc.
For instance, $\bbp(\PPa)$ is the probability for the event that $i_1$ and $j_1$ are connected by one cluster,
and $i_2$ and $j_2$ are connected by another cluster.
Event $\PPb$ differs from event $\PPa$ in the pairing between sites in ${\cal V}_i$ and ${\cal V}_j$, 
i.e., $i_1$ is connected to $j_2$ and  $i_2$ is connected to $j_1$ for $\PPb$.
Within each bulk layer, the probability  for each event is sampled,
and  $\boldsymbol \delta$ is perpendicular to $\bfr$ with $|\boldsymbol \delta|=1$.

For $N=2$ and 3, according to the LCFT theory, 
appropriate linear combinations of the 
probabilities ($\bbp(\PPa)$, $\bbp(\PPb)$, etc.)
give access to the operator content of the underlying 
field theory \cite{Vasseur2012, Couvreur2017}. 
More precisely, each combination corresponds, in the 
continuum limit,
to the two-point function of an operator. This correspondence 
relies on the local ${\cal S}_{N}$ symmetry between the $N$ 
spins of ${\mathcal V}_i$
(or ${\mathcal V}_j$), and the ${\cal S}_Q$ symmetry of the $Q$-state 
Potts model.
Note that ${\cal S}_Q$ is subtly non-trivial, since percolation
is not $Q=1$ but rather $Q \to 1$.
The definitions of observables acting on $N=2$ and $N=3$ spins are recalled below. Each of them corresponds, technically, to
a pair of Young diagrams for ${\cal S}_{N}$ and ${\cal S}_Q$ \cite{ Couvreur2017}.

Consider first observables describing the propagation of $N=2$ clusters. There are two different combinations,
corresponding to the \underline{s}ymmetric and \underline{a}ntisymmetric Young diagrams of ${\cal S}_2$,
\begin{eqnarray}
    P_{2{\rm s}}  = \bbp(\PPa) + \bbp(\PPb) \qquad \mbox{and} \hspace{5mm}
    P_{2{\rm a}}  = \bbp(\PPa) - \bbp(\PPb)  \; .
\label{eq:N2}
\end{eqnarray}
In the continuum limit these correlation functions correspond to the two-point functions of
two operators $\mathcal{O}_{2{\rm s}}$ and $\mathcal{O}_{2{\rm a}}$ respectively.
We use different fonts to distinguish between the probabilities of events which are directly measured in the numerical work, and certain combinations thereof which are found to have particular scaling forms in the continuum limit.
Below, we also use the term observable to describe a two-point function. 
The scaling dimensions of these operators in 2D CFT are available~\cite{Couvreur2017, tan2019observation}. 

For $N=2$, we could also define a logarithmic correlation
function $F$ as~\cite{Vasseur2012}:
\begin{equation} 
F(r)=\frac{  \bP_0(r) + \bP_1(r) - (\bP_{\neq})^2}{P_{2{\rm s}}(r)} \sim \delta \ln (r)\; ,
\label{purelogscaling}
\end{equation}
where $\bP_0 \equiv \bP(\PPzero)$ is the probability that each of the four specified points belongs to a different percolation cluster;
$ \bP_1$ is the probability that the points belong to three different clusters, one of which propagates from one site in ${\mathcal V}_i$
    to another site in $ {\mathcal V}_j$, i.e., $ \bP_1 \equiv \bP(\PPca) + \bP(\PPcb)+\bP(\PPcc)+\bP(\PPcd) $.
    Note that $\bP(\PPzero)$ increases with $r$ and converges to $(\bP_{\neq})^2$ for $r \rightarrow \infty$, where $\bP_{\neq}$ is the
probability that the two points in ${\cal V}_i$ belong to different percolation clusters. 
The composite observable $F$ is expected to behave logarithmically as in Eq.~(2), with $\delta$ a universal factor according to LCFT~\cite{Vasseur2012}.

For $N=3$ clusters, the relevant combinations are
\begin{eqnarray}
    P_{3{\rm s}}   & =& \np \bbp(\PPPa) + \bbp(\PPPb) + \bbp(\PPPc) + \bbp(\PPPd) + \bbp(\PPPe) +\np \bbp(\PPPf) \nonumber \\
    P_{3{\rm m}}  & =& 2 \bbp(\PPPa) + \bbp(\PPPb) + \bbp(\PPPc) - \bbp(\PPPd) - \bbp(\PPPe) - 2 \bbp(\PPPf) \nonumber \\
    P_{3{\rm a}}   & =& \np \bbp(\PPPa)- \bbp(\PPPb)- \bbp(\PPPc)+ \bbp(\PPPd) + \bbp(\PPPe) - \np \bbp(\PPPf) \nonumber  \; ,
\label{eq:N3}
\end{eqnarray}
where $\bbp_{N\circ}$ (with subscript $\circ\! = \! {\rm s}, {\rm m}, {\rm a}$) refers to the \underline{s}ymmetric, \underline{m}ixed and
\underline{a}ntisymmetric Young diagrams of symmetry ${\cal S}_3$.

At criticality, the $N$-cluster correlation functions $\bbp_{N\circ}$  are expected to decay algebraically as $r^{-2 X_{N{\rm o}}}$,
with (a priori) distinct and symmetry-dependent scaling dimensions, i.e., critical exponents $X_{N{\rm o}}$.
It is also predicted that under rotations of the relative angle between the two neighborhoods $ {\mathcal V}_i$ and $ {\mathcal V}_j$,
the amplitudes of the algebraically decaying functions exhibit nontrivial rotational dependences, in accordance with the corresponding conformal spins~\cite{Couvreur2017, tan2019observation}.

The following is the list of sampled quantities,
\begin{itemize}
	\item $N=2$: $P_{2{\rm s}}$, $P_{2{\rm a}}$ and $F$
	\item $N=3$: $P_{3{\rm s}}$,  $P_{3{\rm m}}$ and $P_{3{\rm a}}$ .
\end{itemize}

\section{Simulation and Fitting Method} 
\label{simu}

The simulation of the 4D and 5D bond percolation uses the transfer MC algorithm of Sec.~\ref{algo} and is carried out at the percolation threshold,
which is taken as $p_c ({\rm 4D}) =0.160\, 131\, 22$~\cite{Mertens, Xun2020} and $p_c ({\rm 5D}) = 0.118\, 171\, 45$~\cite{Mertens}.
In 4D, we take system sizes $L=$4, 6, 8, 10, 12, 14, 16, 18, 20, 24, 32, 40, 48, 64,  96, 128, 192, 256, 384, 512. 
For each half-infinite cylinder, more than $2\times 10^9$ surface layers are generated for each $L \le 32$, 
and at least $4\times 10^7$ layers for each $L > 64$. 
In 5D, we take $L =$ 4, 6, 8, 10, 12, 14, 16, 20, 24, 28, 32, 40, 48, 64, 80, 96, 128,
and for each half-infinite cylinder,  generate more than $5\times 10^8$ surface layers for each $L \le 24$ and
at least $10^7$ layers for each $L > 24$. 
Initial simulations of $10L$ layers for each half-infinite cylinder are discarded before measurements are taken.
In total, about $2\times 10^6$ CPU hours $\approx 228$ CPU years are used.

Measurements are taken within each of the two finite bulk layers of $L^{d-1}$ lattice sites.
According to the finite-size scaling theory~\cite{cardy_book}, we expect that at criticality, 
the $N$-cluster correlation functions of distance $r$ behave as
 \begin{equation}
     P_{N \rm o}(r,L)\sim r^{-2X_{N\rm o}} \widetilde{P}_{N \rm o}(r/L) \; ,
 \label{eq:FSSc}
 \end{equation}
 where $\widetilde{P}_{N \rm o}$ is a universal scaling function and $X_{N \rm o}$ is an $N$-cluster exponent
 (with subscript $\circ \in$ \{s, a\} for $N=2$ and $\circ\in$\{s, m, a\} for $N=3$).
 For simplicity, we set $r=L/2$ so that $P_{N \rm o}(r,L)$ only depends on the linear size $L$ as 
  \begin{equation}
    P_{N \rm o}(L)\sim L^{-2X_{N \rm o}} .
 \end{equation}
 Note that one can  in principle take $r=aL$ with $0 < a <1/2$ a constant. 
 Nevertheless, while a smaller value of $a$ would enhance the amplitude of $P_{N \rm o}$, 
 stronger finite-distance corrections would also occur. 
 Meanwhile, the choice of $r=L/2$ benefits from the fact that the $N$-cluster correlators have equal contributions from both 
 directions of connection in the periodic system.

We expect that the logarithmic function $F(r=L/2)$  diverges logarithmically as $F(L) \asymp \tilde{\delta} \ln (L)$. 
Nevertheless, since  the correlators at distance $r=L/2$ in Eq.~(\ref{purelogscaling}) are unavoidably affected by finite system sizes, 
as reflected by the finite-size scaling form in Eq.~(\ref{eq:FSSc}), it is not clear whether the amplitude $\tilde{\delta}$ 
is universal and equal
to the coefficient $\delta$ obtained by studying the $r$-dependence.
Therefore, we carry out further extensive simulations to measure the universal amplitude $\delta$ in a new procedure, in which 
the system sizes are fixed to $L=384$ in 4D and $L=96$ in 5D. 
We take distances $r=3,4,5,6,7,8,10,13,16$ in 4D and  $r=3,4,5,6,7,8$ in 5D. 
For each half-infinite cylinder, more than $3\times 10^7$ ($6\times 10^7$) surface layers are generated in 4D (5D). 
The extra simulations alone took about $10^6$ CPU hours $\approx 114$ CPU years.

According to the least-squares criterion, we fit the MC data of the $N$-cluster correlation functions to 
\begin{equation}
    P_{N \rm o}(L) = L^{-2X_{N \rm o}}\left(a +b L^{y_1}\right) \; ,
\label{eq_fit}
\end{equation}
the logarithmic correlation function $F(r=L/2)$ to
\begin{equation}
    F(L) = a +\tilde{\delta} \ln(L) \; ,
\label{eq_fit_log_L}
\end{equation}
and logarithmic correlation function $F(r)$ to
\begin{equation}
    F(r) = a +\delta \ln(r),
\label{eq_fit_log}
\end{equation}
where $a $ is a constant and $b L^{y_1}$ accounts for the leading finite-size correction term with exponent $y_1<0$. 

As a precaution against other correction-to-scaling terms which we fail to include in the fitting ansatz, we impose
a lower cutoff $L\ge L_{{\rm m}}$ ($r\ge r_{\rm m}$) on the data points admitted in the fits, and systematically study the effect on the $\chi^2$ value when increasing $L_{\rm m}$ ($r_{\rm m}$). Generally, we prefer fits corresponding to the smallest $L_{\rm m}$ ($r_{\rm m}$) for which the goodness of fit is reasonable and subsequent increase in $L_{\rm m}$ ($r_{\rm m}$) does not cause the $\chi^2$ value to drop by vastly more than one per degree of freedom. In practice, by `reasonable' we mean that $\chi^2/{\rm DF}\lesssim 1$, where `DF' is the number of degrees of freedom.

The error of our estimates consists of two parts, the statistical error and the systematic error. 
The statistical error is the error of MC simulations due to the randomness of the sampling procedure. 
All the observables have a statistical error, which will enter into the fitting results. 
The errors listed in the tables for each individual fit are all statistical errors. 
The systematic error is due to finite-size corrections. 
To account for it, we perform fits with different values of $L_{{\rm m}}$ and $y_{{\rm 1}}$. 
The confidence interval of our final estimates are set to be the union of confidence intervals in all individual fits with different fit conditions.

\section{Results}
\label{result}

\begin{figure}
	\includegraphics[width=\linewidth]{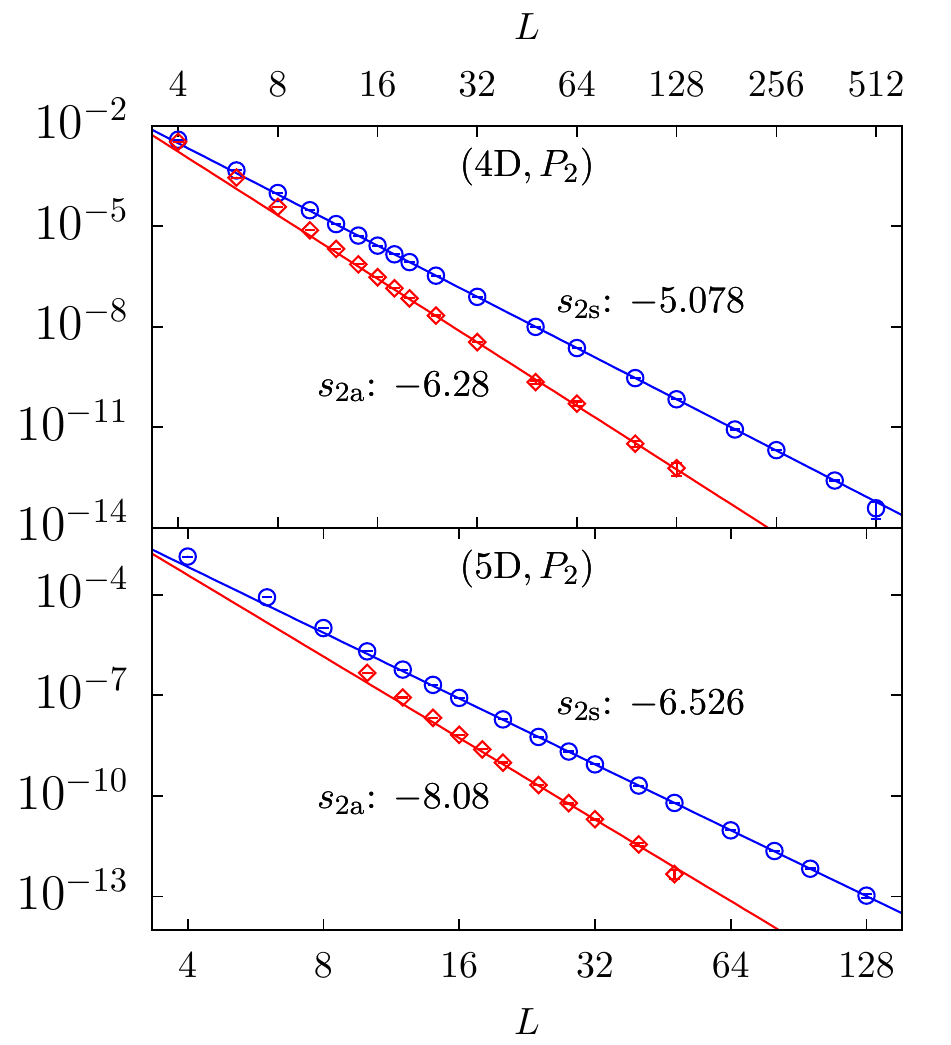}
    \caption{ Log-log plot of $P_{\rm 2\circ} (\circ={\rm s, a})$ versus  $L$ for  4D and 5D. 
		The straight lines with slope $s$ come from the least-squares fits.}
	\label{fig_p2sa}
\end{figure}

\subsection{$N=2$ correlation functions}

The $N=2$ correlation functions scale as $P_{\rm 2s}\sim L^{-2X_{\rm 2s}}$ and $P_{\rm 2a} \sim L^{-2X_{\rm 2a}}$. 
The MC data for these correlation functions are shown in Fig.~\ref{fig_p2sa}.
We fit the data for $P_{\rm 2s}$ and $P_{\rm 2a}$ to Eq.(\ref{eq_fit}) 
and the fitting results are shown in Table~\ref{tab_p2sa}.

\begin{table}[htb]
    \centering\small
    \begin{tabular}{|l|l|l|l|l|l|l|}
        \hline
                     &$d$ & $X$         & $a $   & $b$  &  $y_1$   & $L_{\rm m}/DF/\chi^2$\\
        \hline
\multirow{12}{0.5cm}{$P_{\rm 2s}$} & \multirow{6}{0.2cm}{4}   & 2.5382(5)   &3.40(1)  & 3927(2000) & -4.3(2)  & 12/11/15 \\
                     &    & 2.5383(8)   &3.40(2)  & 1800(4000) & -4(1)    & 16/9/11 \\
                     &    & 2.5382(4)   &3.40(1)  & 1400(110)  & -4       & 16/10/11 \\
                     &    & 2.5388(6)   &3.41(2)  &  900(400)  & -4       & 20/8/8 \\
                     &    & 2.5395(4)   &3.43(1)  &  /         &   /      & 24/8/9 \\
                     &    & 2.5388(8)   &3.41(2)  &  /         &   /      & 32/7/8 \\
        \cline{2-7}
                 & \multirow{6}{0.2cm}{5}  & 3.265(1)   &5.89(5)  & 1386(140) & -3.10(5)  & 10/11/14 \\
                     &    & 3.265(2)   &5.86(8)   & 1266(330)   & -3.06(11)  & 12/10/14 \\
                     &    & 3.262(3)   &5.76(13)  & 735(415)    & -2.8(3)  & 14/9/13 \\
                     &    & 3.2633(7)   &5.81(2)  & 1110(10) & -3  & 12/10/13 \\
                     &    & 3.264(1)    &5.82(4)  & 1099(22)   & -3  & 14/9/13 \\
                     &    & 3.263(2)    &5.80(6)  & 1120(44)   & -3  & 16/8/12 \\
        \cline{1-7}
\multirow{15}{0.5cm}{$P_{\rm 2a}$} & \multirow{8}{0.2cm}{4}    & 3.17(2)   &11.8(13)  & 10400(9000)  & -3.3(4)  & 12/7/11 \\
                     &    & 3.13(5)   &9(3)  & 800(1400)  & -2.3(7)  & 12/6/8 \\
                     &    & 3.12(1)   &7.9(7)  & 440(8)  & -2  & 14/7/9 \\
                     &    & 3.13(2)   &8.9(10) & 420(25) & -2  & 16/6/7 \\
                     &    & 3.10(2)   &7(1)    & 480(30) & -2  & 18/5/5 \\
                     &    & 3.168(7)   &11.7(6)  & 4400(140) & -3  & 14/7/8 \\
                     &    & 3.17(1)   &12(1)  & 4400(400) & -3  & 16/6/8 \\
                     &    & 3.14(1)   &10(1)  & 5880(500) & -3  & 18/5/4 \\
        \cline{2-7}
                     & \multirow{7}{0.2cm}{5}   & 4.02(3) & 26(6) &22600(7300) & -2.9(1) & 10/6/6 \\
                     &    & 4.03(7)   &28(14)  & 27000(33000) & -3.0(5)   & 12/5/5 \\
                     &    & 4.05(2)   &31(3)  & 28600(1300) & -3   & 10/7/6 \\
                     &    & 4.03(2)   &28(4)  & 27900(1800) & -3   & 12/6/5 \\
                     &    & 4.06(3)   &32(7)  & 28600(1800) & -3   & 14/5/5 \\
                     &    & 3.99(5)   &20(6)  & 7600(900) & -5/2   & 14/5/5 \\
                     &    & 4.03(7)   &25(14) & 7700(740) & -5/2   & 16/4/4 \\
        \hline
    \end{tabular}
    \caption{Fits for $N=2$ correlation functions.}
    \label{tab_p2sa}
\end{table}

The $P_{\rm 2s}$ data are well described by Eq.(\ref{eq_fit}) with $y_1$ fixed to $-4$ and $-3$ in 4D and 5D respectively. 
In 4D,  the data for $L_{\rm m}\geq 24$ can be fitted without the finite-size correction term--i.e., $b=0$.
From these fits we take our final estimate to be $X_{\rm 2s}({\rm 4D}) = 2.5390(12)$ and $X_{\rm 2s}({\rm 5D}) =3.263(2)$.

The exponent $X_{\rm 2s}$  is also called  the two-arm exponent~\cite{Smirnov2001, Aizenman1999}. 
For the case of percolation, it is further related to the thermal renormalization exponent $y_{\rm t}=1/\nu$ or the red-bond exponent $y_{\rm red}$ 
as   $X_{\rm 2s}=d-y_{\rm t}=d-y_{\rm red}$~\cite{Stanley77, Xu2014a}. 
Thus, from the results for $X_{\rm 2s}$ we obtain
$y_{\rm t}({\rm 4D})=1.4610(12)$ and $y_{\rm t}({\rm 5D})=1.737(2)$.
In comparison with the most recent results $y_{\rm t}({\rm 4D})=1.459(6)$ and $y_{\rm t}({\rm 5D})=1.747(5)$ in Ref.~\cite{Koza16},
our estimates have a much higher precision.

\begin{figure}
    \includegraphics[width=0.85\linewidth]{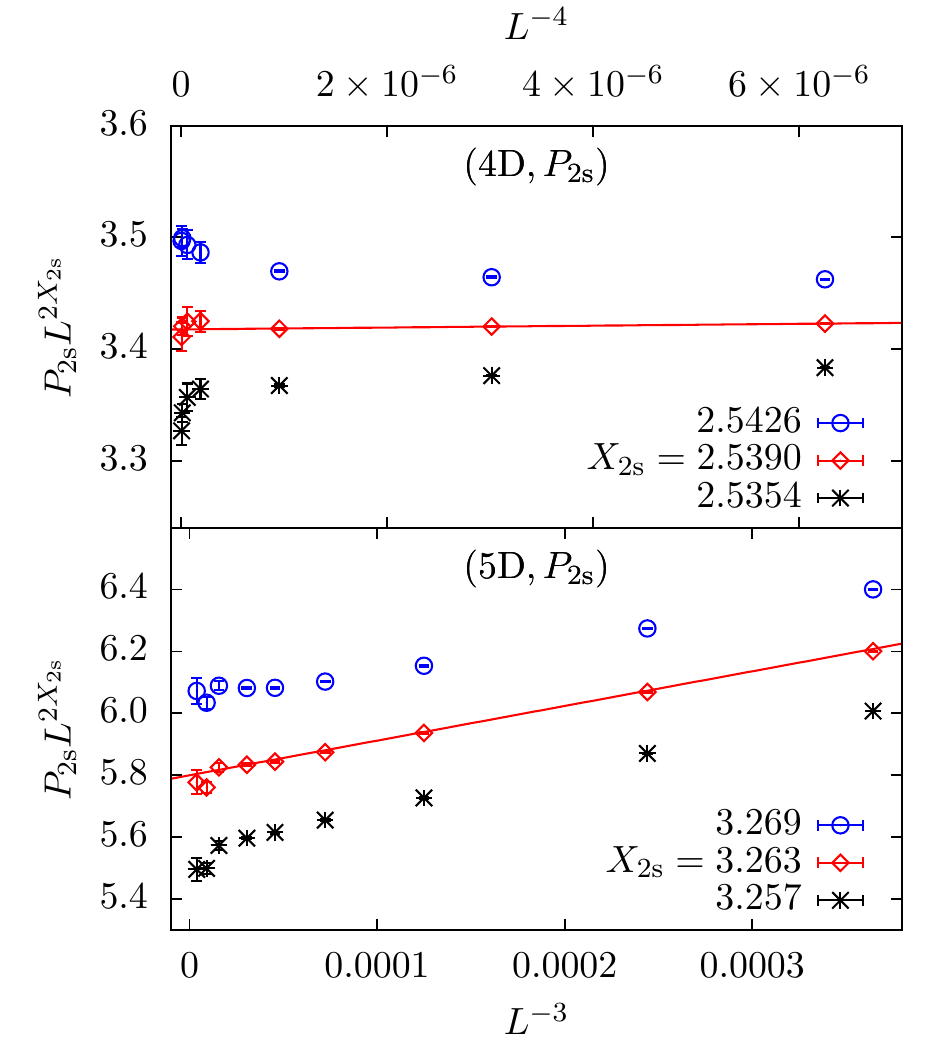}
	\caption{ Plot of $P_{\rm 2s} L^{2X_{\rm 2s}}$ versus $L^{-4}$ ($L^{-3}$) for 4D (5D), 
		illustrating the reliability of our estimate $X_{\rm 2s}({\rm 4D}) = 2.5390(12)$ and  $X_{\rm 2s}({\rm 5D})=3.263(2)$.
		The straight lines are obtained from the fits.}
	\label{fig_p2s}
\end{figure}

To examine the reliability of our final results, 
we plot  $P_{\rm 2s} L^{2X_{\rm 2s}}$ vs $L^{-4}$ ($L^{-3}$) for 4D (5D) in Fig.~\ref{fig_p2s}.
Three different values of $X_{\rm 2s}$ are used, corresponding to our final quoted value, 
as well as those with three standard deviations $3\sigma$ away.
Using $y_{\rm t}({\rm 4D})=1.4610$ and $y_{\rm t}({\rm 5D})=1.737$, approximately straight lines are produced 
for both 4D and 5D, and in the $L \rightarrow \infty$ limit, the  $P_{\rm 2s} L^{2X_{\rm 2s}}$ values quickly converge to some constants.
It can be seen that finite-size corrections are rather minor, particularly in 4D.
By contrast, when using the values away by $3\sigma$, the $P_{\rm 2s} L^{2X_{\rm 2s}}$ data significantly bend upward or downward,
illustrating the robustness of our results.  
In 5D, the black star data points correspond to $y_{\rm t}=1.743$. 
If $y_{\rm t}=1.747$ ($X_{\rm 2s}=d-y_{\rm t}=3.253$), $5\sigma$ away from our estimate, were used, 
the bending-down curvature would be even more severe. 
Thus, even though our estimate $y_{\rm t}=1.737(2)$ and result $y_{\rm t}=1.747(5)$ in Ref.~\cite{Koza16}
are basically consistent, the central value $y_{\rm t}=1.747$ is nearly excluded for 5D.

\begin{figure}
    \includegraphics[width=0.85\linewidth]{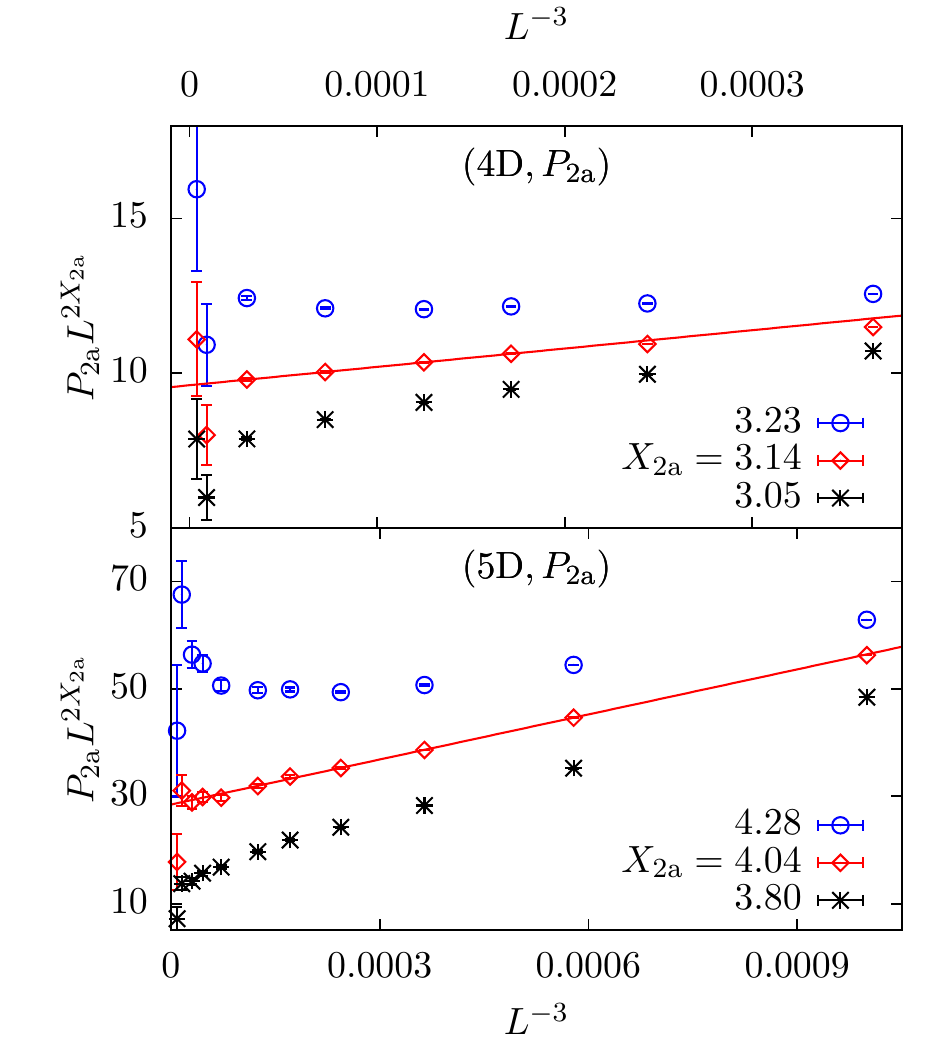}
	\caption{ Plot of $P_{\rm 2a} L^{2X_{\rm 2a}}$ versus $L^{-3}$ for 4D and 5D, 
        illustrating our estimate $X_{\rm 2a}({\rm 4D}) = 3.14(3)$ and  $X_{\rm 2a}({\rm 5D})=4.04(8)$.
		The straight lines are obtained from the fits.}
	\label{fig_p2a}
\end{figure}

Compared to $P_{\rm 2s}$, the antisymmetric correlation function $P_{\rm 2a}(L)$ decays much faster as a function of distance $r$,  
 because of the cancellation between the connectivity probabilities $\bbp(\PPa)$ and  $\bbp(\PPb)$;
the $P_{\rm 2a}(L)$ data are also less precise. 
The fitting results for  $P_{\rm 2a}$ are also shown in Table~\ref{tab_p2sa}. 
In 4D, leaving $y_1$ as a free parameter in the fit shows that $y_1$ is around 2 and 3 with large uncertainties. 
From the fitting results with $y_1$ fixed to 2 and 3, we obtain $X_{\rm 2a}({\rm 4D})=3.14(3)$. 
For 5D, our final estimate is $X_{\rm 2a}({\rm 5D}) =4.04(8)$.
Following the same trick as for $P_{\rm 2s}$, the reliability of our results is examined 
in Fig.~\ref{fig_p2a},  which plots $P_{\rm 2a}L^{2X_{\rm 2a}}$ versus $L^{-3}$ for 4D and 5D
using three different values of $X_{\rm 2a}$. 

\subsection{Logarithmic correlation functions}

\begin{figure}
    \includegraphics[width=0.85\linewidth]{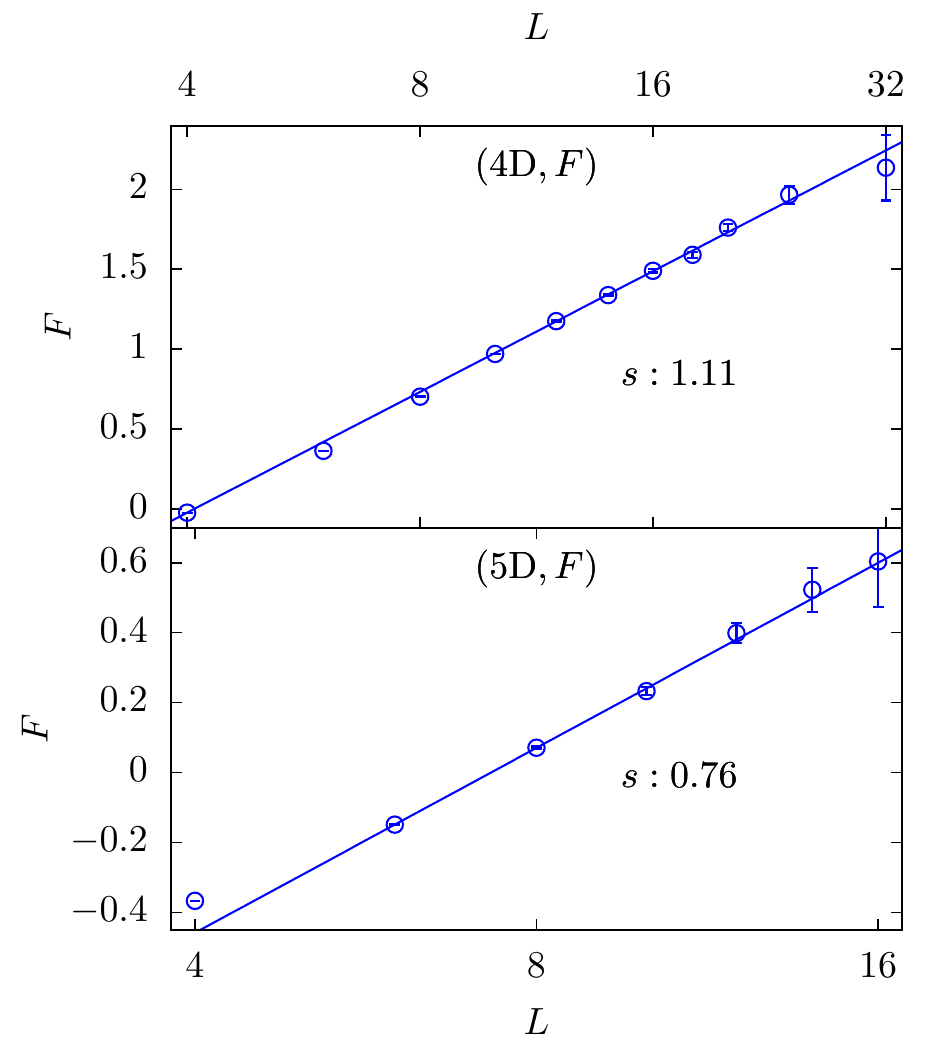}
	\caption{Semi-log plot of the logarithmic correlation $F(L)$ for 4D and 5D. 
	The logarithmic behavior is clearly shown, especially in 4D. 
    The slopes of the straight lines  are respectively 
    $\tilde{\delta}({\rm 4D}) = 1.11(5)$ and $\tilde{\delta}({\rm 5D}) = 0.76(4)$.}
	\label{fig_fr}
\end{figure}

\begin{figure}
    \includegraphics[width=0.85\linewidth]{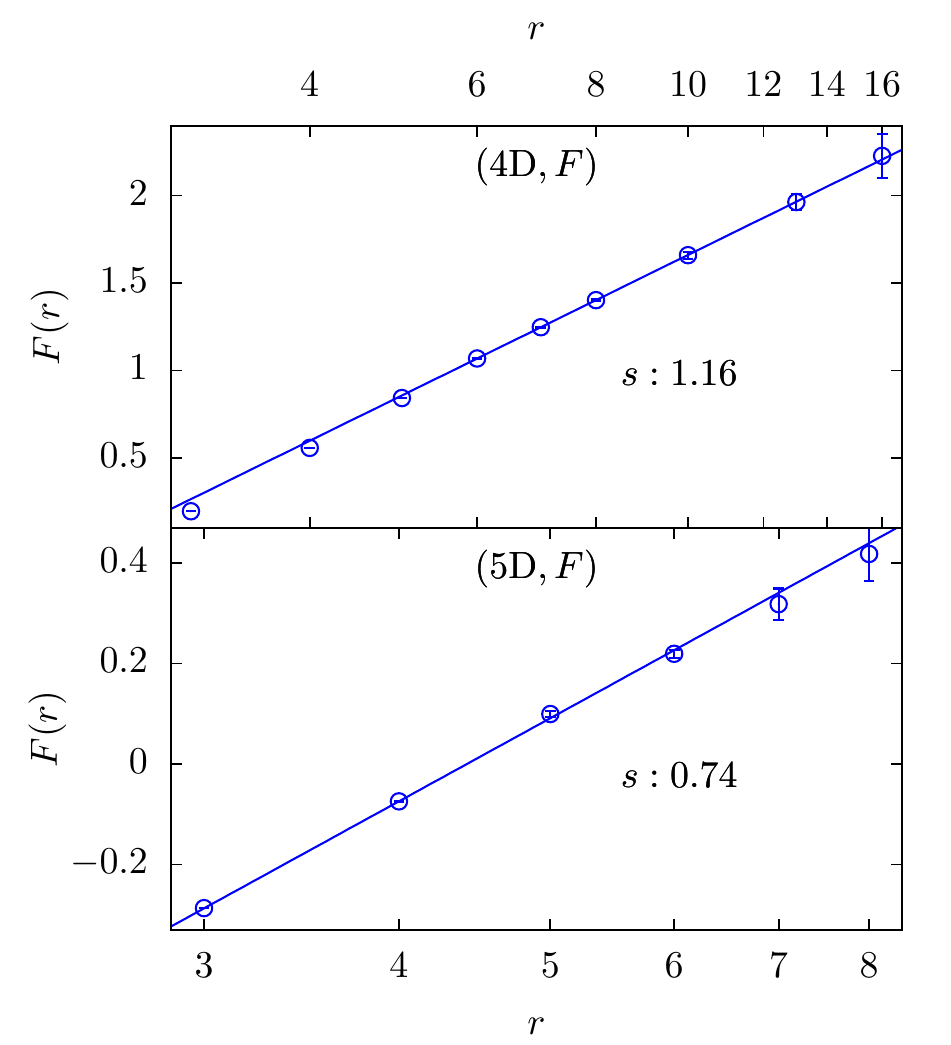}
	\caption{Semi-log plot of the logarithmic correlation $F(r)$ for 4D and 5D. 
    The system sizes are $L=384$ in 4D and $L=96$ in 5D.
	The logarithmic behavior is clearly shown, especially in 4D. 
    The slopes of the straight lines are universal, and their values are respectively 
    $\delta({\rm 4D}) = 1.16(1)$ and $\delta({\rm 5D}) = 0.74(6)$.}
	\label{fig_fr_r}
\end{figure}

In the definition of the composite logarithmic correlation function $F$ by Eq.~({\ref{purelogscaling}}), 
there exist cancellations among various connectivity probabilities in the numerator, as well as 
cancellations of the algebraic decay between the numerator and the denominator, 
leading to the logarithmic divergence as a function of distance $r$.
The exact structure in Eq.~({\ref{purelogscaling}}) is derived directly in the thermodynamic limit. 
It is not a priori clear that the cancellations in $F(r,L)$,  calculated at half linear size $r=L/2$, 
are the same to ensure a logarithmic finite-size dependence $F(r=L/2) \sim \ln L$. 

\begin{table}[htb]
    \centering\small
    \begin{tabular}{|l|l|l|l|l|}
        \hline
                     &$d$  & $\tilde{\delta}$   & $a$    & $L_{\rm m}/DF/\chi^2$\\
        \hline
        \multirow{5}{0.8cm}{$F(L)$} & \multirow{3}{0.2cm}{4}   & 1.11(1)  &-1.5(3)   & 10/6/6 \\
                     &    & 1.09(2)   &-1.54(6)        & 12/5/5 \\
                     &    & 1.12(5)   &-1.6(1)         & 14/4/4 \\
        \cline{2-5}
                 & \multirow{2}{0.2cm}{5}  & 0.763(5)  &-1.52(1)  &6/4/2 \\
                     &    & 0.76(2)   &-1.51(5)                   &8/3/2 \\

        \hline
                     &$d$  & $\delta$   & $a$    & $r_{\rm m}/DF/\chi^2$\\
        \hline
        \multirow{5}{0.8cm}{$F(r)$} & \multirow{3}{0.2cm}{4}   & 1.159(2)  &-1.010(3)   & 6/1/4 \\
                     &    & 1.160(4)   &-1.01(1)        & 7/1/3 \\
                     &    & 1.157(8)   &-1.00(2)        & 8/1/2 \\
        \cline{2-5}
                 & \multirow{2}{0.2cm}{5}  & 0.74(2)  &-1.10(2)  &4/4/3 \\
                     &    & 0.67(3)   &-0.97(5)        & 6/1/1 \\
        \hline
    \end{tabular}
    \caption{Fits for logarithmic correlation function $F(L)$ and $F(r)$.}
    \label{tab_fr}
\end{table}

The data for $F(L)$ are shown in Fig.~\ref{fig_fr}, where the largest system size is rather limited---$L({\rm 4D})=32$ and $L({\rm 5D})=16$. 
Due to the cancellations, it is computationally expensive to have accurate MC data for larger $L$.
Nevertheless, the approximately straight lines 
in the semi-log plot (Fig.~\ref{fig_fr}) clearly show the logarithmic diverging behavior $F \sim \ln L$.
According to the least-squares criterion, the data are fitted to Eq.~(\ref{eq_fit_log_L}) and the results are shown in Table~\ref{tab_fr}.
From the fitting results, we take our final estimate of the amplitude $\tilde{\delta}$ to be 
$\tilde{\delta}({\rm 4D}) = 1.11(5)$ and $\tilde{\delta}({\rm 5D}) =0.76(4)$.

The data for $F(r)$ are shown in Fig.~\ref{fig_fr_r}, which clearly confirm the logarithmic diverging behavior $F \sim \delta \ln r$.
They are fitted to Eq.~(\ref{eq_fit_log}) and the results are shown in Table~\ref{tab_fr}.
We take our final estimates to be $\delta({\rm 4D}) = 1.16(1)$ and $\delta({\rm 5D}) =0.74(6)$,
consistent with the results for $\tilde{\delta}$. 
In 4D, $\delta$ has a significantly smaller error than $\tilde{\delta}$.

To further check whether the two amplitudes always agree, we have also performed extensive simulations to measure $\tilde{\delta}$ in 2D.
In previous work we found $\delta = 1.12(3)$~\cite{tan2019observation}, in good agreement 
with the analytical result $\delta({\rm 2D)} = 2 \sqrt{3} / \pi$~\cite{Vasseur2012}.
We now obtain $\tilde{\delta} = 1.03(1)$ in 2D, indicating that the two amplitudes do differ in general. 
By contrast, the difference in 4D is at best marginally
discernible, and in 5D not at all, to within the numerical precision of the data shown in Table~\ref{tab_fr}.

\begin{figure}
	\includegraphics[width=0.85\linewidth]{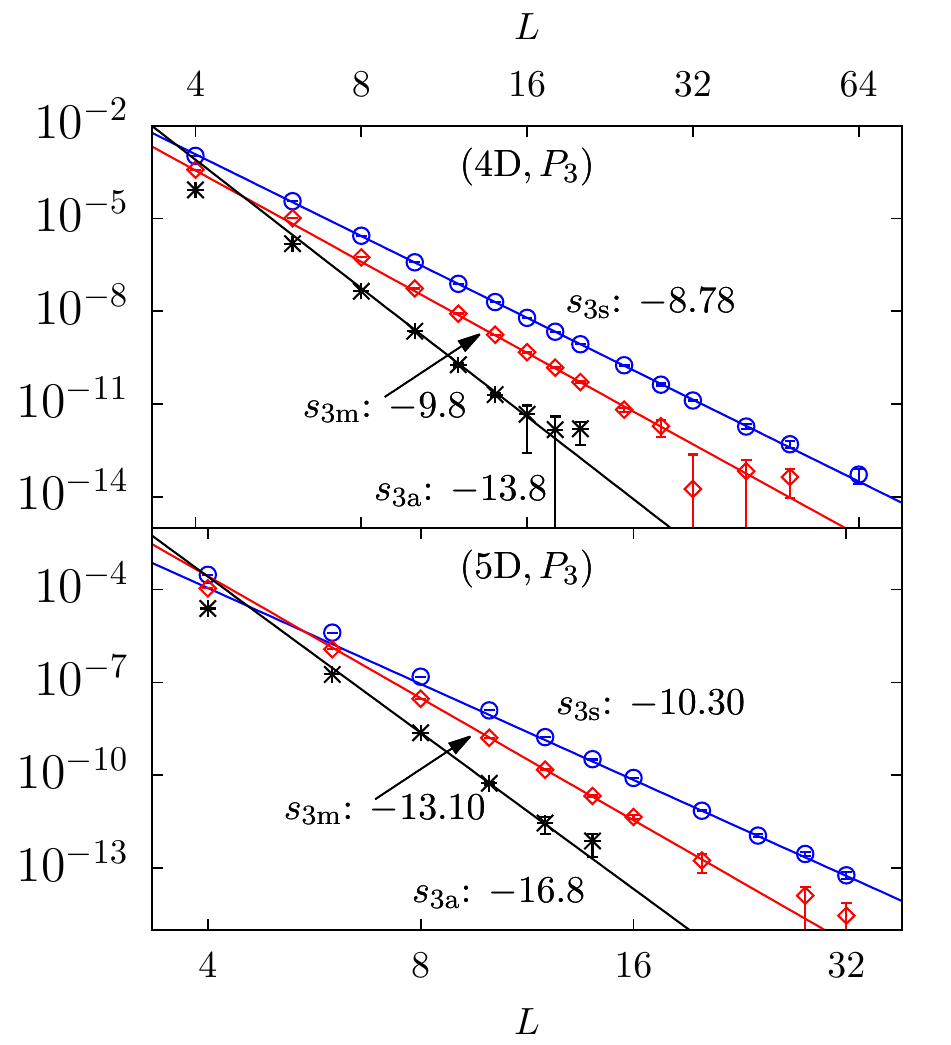}
    \caption{ Log-log plot of $P_{\rm 3\circ}(\circ={\rm s, m, a})$ versus $L$ for  4D and 5D. 
    For clarity, the $P_{\rm 3s}$ ($P_{\rm 3m}$) data have been multiplied by a factor 10 (2).
		The straight lines with slope $s$ come from the least-squares fits.}
	\label{fig_p3sam}
\end{figure}

\subsection{$N=3$ correlation functions }
The $N=3$ correlation functions include $P_{\rm 3s}$, $P_{\rm 3m}$ and $P_{\rm 3a}$, and the data are shown in Fig.~\ref{fig_p3sam}.
As $N$ increases, the probability that $N$ distinct clusters propagate from a small neighborhood to another one far away drops quickly.
The $N=3$ correlation functions decay very rapidly as $L$ increases, particularly the antisymmetric correlator $P_{\rm 3a}$.
Because of cancellations of various connectivities in Eq.~(\ref{eq:N3}), 
the scaling of $P_{\rm 3m}$ is effectively the sub-leading behavior of 3-cluster propagating probabilities,
and that of $P_{\rm 3a}$ corresponds to the sub-sub-leading behavior.
As a result, it is very challenging to obtain meaningful data with reliable statistical errors for large $N$ and/or large $L$.

\begin{table}[htb]
    \centering\small
    \begin{tabular}{|l|l|l|l|l|l|l|}
        \hline
                    &$d$ & $X$   & $a$   & $b$  &  $y_1$   & $L_{\rm m}/DF/\chi^2$\\
        \hline
\multirow{8}{0.5cm}{$P_{\rm 3s}$} & \multirow{4}{0.2cm}{4} & 4.365(6)   &19.7(7)  & 780(70) & -3  & 8/10/8 \\
                     &    & 4.384(11)   &22(1)    & 300(270) & -3  & 10/9/6 \\
                     &    & 4.396(3)    &23.6(3)  & /        &  /  & 10/10/6 \\
                     &    & 4.390(6)    &22.9(7)  & /        &  /  & 12/9/6 \\
        \cline{2-7}
        & \multirow{4}{0.2cm}{5}   & 5.12(6)    &15(5)   & 2200(260)  & -5/2  & 8/6/6 \\
                           &       & 5.12(10)   &15(9)   & 2200(330)  & -5/2  & 10/5/5 \\
                           &       & 5.19(4)    &25(5)   & 6400(400)  & -3    & 8/6/6 \\
                           &       & 5.16(7)    &20(9)   & 6500(600)  & -3    & 10/5/5 \\
        \cline{1-7}
        \multirow{4}{0.5cm}{$P_{\rm 3m}$} & \multirow{2}{0.2cm}{4} & 4.94(6)   &172(59)  & $30(2)\times10^4$  & -4  & 10/5/5 \\
                           &       & 4.87(12)   &121(80)  & $30(6)\times10^4$  & -4  & 12/4/5 \\

        \cline{2-7}
        & \multirow{2}{0.2cm}{5} & 6.54(1)   &9900(300)   & $15.0(4)\times10^5$  & -4  & 4/5/5 \\
        &                        & 6.55(3)   &10860(1600) & $18(6)\times10^5$    & -4  & 6/4/1 \\
        \cline{1-7}
        \multirow{4}{0.5cm}{$P_{\rm 3a}$} & \multirow{2}{0.2cm}{4} & 6.95(5)   &$2.1(5)\times 10^{5}$  & $-23(6)\times10^6$  & -3  & 6/5/3 \\
          &    & 6.85(30)   &$1.2(17)\times 10^{5}$  & $-8(20)\times10^6$  & -3  & 8/4/3 \\
        \cline{2-7}
         & \multirow{2}{0.2cm}{5} & 7.6(1)   &$1.1(1)\times 10^{5}$  & /   & /  & 6/6/20 \\
                 &                & 8.4(3)   &$3(3)\times 10^{6}$  & /   & /  & 8/5/6 \\
        \hline
    \end{tabular}
    \caption{Fits for $N=3$ correlation functions.}
    \label{tab_p3sam}
\end{table}

\begin{figure}
    \includegraphics[width=0.85\linewidth]{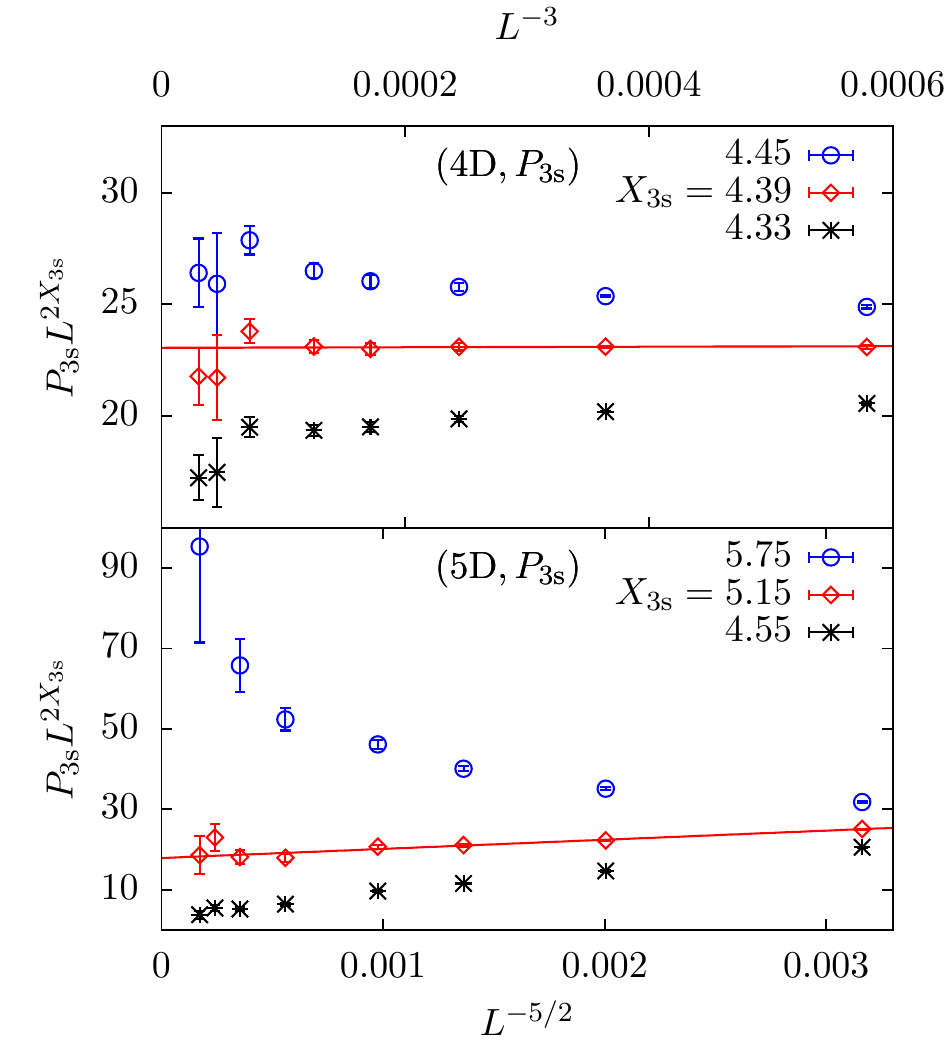}
	\caption{ Plot of $P_{\rm 3s} L^{2X_{\rm 3s}}$ versus $L^{-3}$ ($L^{-5/2}$) for  4D (5D), 
        illustrating our estimate $X_{\rm 3s}({\rm 4D}) = 4.39(2)$ and  $X_{\rm 3s}({\rm 5D})=5.15(20)$.
		The straight lines are obtained from the fits.}
    \label{fig_p3s}
\end{figure}

For the symmetric correlator $P_{\rm 3s}$, we obtain data with maximum $L=64$ for 4D and $32$ for 5D, 
with magnitudes in ${\cal O}(10^{-14})$ and relative errors about $20\%$. 
The fitting results are shown in Table~\ref{tab_p3sam}.
In 4D, $P_{\rm 3s}$ data for $L \geq 8$ can be well described with correction exponent $y_1=-3$
and the correction amplitude vanishes, $b=0$, for $L_{\rm m} \geq 10$.
In 5D, $P_{\rm 3s}$ data can be described with correction exponent $y_1$ around $-3$.
The fits with $y_1=-3$ or $-5/2$ give consistent results $L_{\rm m}\geq 8$ .
From these fits we take our final estimate to be $X_{\rm 3s}$(4D) = 4.39(2) and $X_{\rm 3s}$(5D) =5.15(20),
of which the reliability is tested in Fig.~\ref{fig_p3s}.

\begin{figure}
    \includegraphics[width=0.85\linewidth]{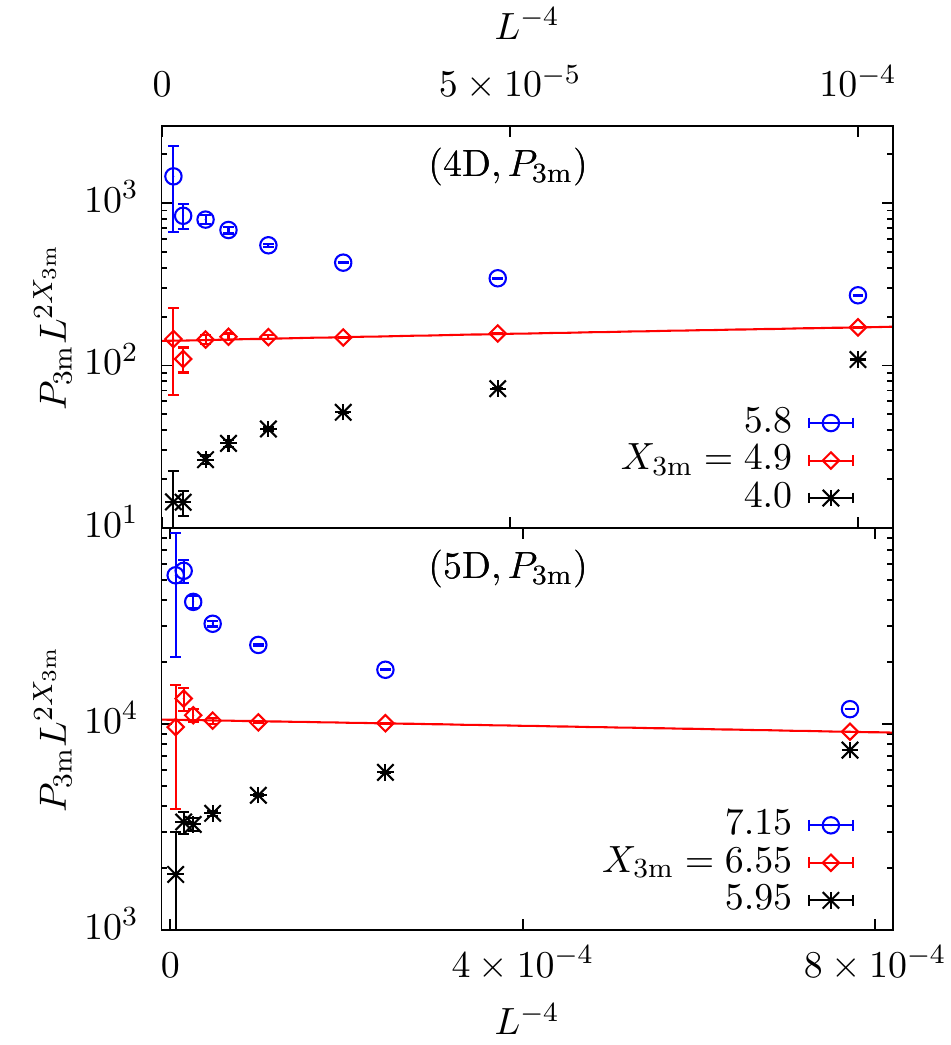}
		\caption{ Plot of $P_{\rm 3m} L^{2X_{\rm 3m}}$ versus $L^{-4}$ for 4D and 5D, 
		illustrating our estimate $X_{\rm 3m}({\rm 4D}) = 4.9(3)$ and  $X_{\rm 3m}({\rm 5D})=6.55(20)$.
        The lines are obtained from the fits. The log scale on the $y$-axis is taken because of the large $y$ range.}
    \label{fig_p3m}
\end{figure}

For $P_{\rm 3m}$, we have data with maximum $L=28$ for 4D and $16$ for 5D, 
which has a magnitude in ${\cal O}(10^{-12})$ and relative errors about $20\%$. 
The fitting results by Eq.~(\ref{eq_fit}) are shown in Table~\ref{tab_p3sam}.
In both 4D and 5D, $P_{\rm 3m}$ data can be described with correction exponent $y_1=-4$. 
From these fits we take our final estimate to be $X_{\rm 3m}$(4D) = 4.9(3) and $X_{\rm 3m}$(5D) =6.55(20).
In Fig.~\ref{fig_p3m}, we plot $P_{\rm 3m}L^{2X_{\rm 3m}}$ vs $L^{-4}$ for 4D and 5D
using three different values of $X_{\rm 3m}$.

\begin{figure}
    \includegraphics[width=0.85\linewidth]{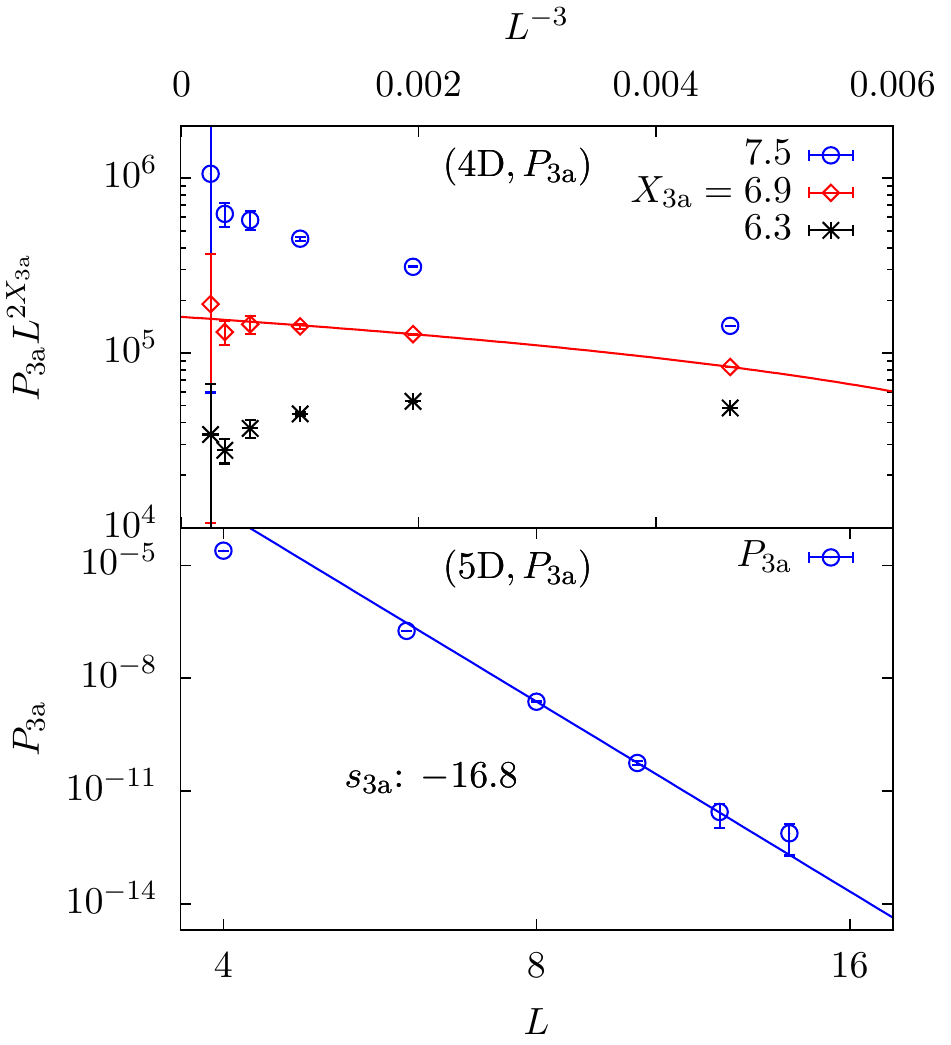}
	\caption{ Upper panel:  $P_{\rm 3a} L^{2X_{\rm 3a}}$ versus $L^{-3}$ for  4D, 
		illustrating our estimate $X_{\rm 3a}({\rm 4D}) = 6.9(2)$.
        The lines are obtained from the fits. 
        Note that the $y$ axis is on a log scale because of its large range, and thus the bending curvatures are significant.
	    Lower panel: log-log plot of $P_{\rm 3a} $ versus $L$ for  5D. 
		The straight line with slope $s$ comes from the least-squares fits.}
    \label{fig_p3a}
\end{figure}

It is very difficult to obtain a reliable estimate of the exponent $ X_{\rm 3a}$ for the antisymmetric correlator $P_{\rm 3a}$, 
which decays extremely rapidly as $L$ increases. 
For instance, $P_{\rm 3a}$ in 5D decays algebraically with exponent $2X_{\rm 3a}\approx 16.0$, leading to a tiny value $\leq 10^{-14}$ already for $L=16$.
In addition, as a residual effect that survives during cancellations of various  3-cluster connectivities, 
$P_{\rm 3a}$ has the same statistical variance as the symmetric correlator $P_{\rm 3s}$,
making it hard to reduce the statistical error of $P_{\rm 3a}$.
We have the $P_{\rm 3a}$ data only up to $L=14$ for 4D and $10$ for 5D, which has a magnitude in ${\cal O}(10^{-11})$ 
and relative errors about  10\%. The 5D $L=12$ data point has a higher relative error as shown in Fig.~\ref{fig_p3a}. 
The fitting results by Eq.~(\ref{eq_fit}) are shown in Table~\ref{tab_p3sam}.
From these fits we take our final estimate to be $X_{\rm 3a}$(4D) = 6.9(2) and $X_{\rm 3a}({\rm 5D}) = 8.4$.
In 5D, since the number of data points is too small,  we are not able to give a robust error bar for exponent $X_{\rm 3a}$.
In Fig.~\ref{fig_p3a}, we plot $P_{\rm 3a}L^{2X_{\rm 3a}}$ vs $L^{-3}$
for 4D using three different values of $X_{\rm 3a}$ in the upper panel, and $P_{\rm 3a}$ vs $L$ directly for 5D  in the lower panel.

\section{Discussion}
\label{discc}

 \begin{table} 
 \begin{ruledtabular}
 \begin{tabular}{llllll}
 $d$                 & 2              & 3                         & 4                     &5                        & $\geq 6$           \\
 \hline 
 $X_{\rm 1}$    & $5/48$     & $0.47707(10)$   & $0.9554(7)$     & $1.4740(14)$    & 2    \\
 \hline
 $X_{\rm 2s}$   & $5/4$       & $1.8587(15)$         & $2.5390(12)$   & $3.263(2)$        & 4    \\
 $X_{\rm 2a}$   & $23/16$   & $2.262(10)$       & $3.14(3)$         & $4.04(8)$        & /     \\
 \hline
 $X_{\rm 3s}$   & $35/12$   & $3.605(8)$         & $4.39(2)$         & $5.15(20)$        & 6     \\
 $X_{\rm 3m}$  & $3$          & $3.93(4)$           & $4.9(3)$           & $6.55(20)$        & /     \\
 $X_{\rm 3a}$   & $11/3$     & $5.2(2)$              & $6.9(2)$          & $8.4$                & /      \\
 \hline
 $\delta$           & $2\sqrt{3}/\pi$ & $1.52(3)$     &  $1.16(1)$       &  $0.74(6)$       & /     
 \end{tabular}
\end{ruledtabular}
    \caption{\label{tab:exponent} $N$-cluster exponents $X$ ($1\leq N\leq 3$) and the universal logarithmic amplitude $\delta$ 
    as a function of spatial dimension $d$.
    The  exponent $X_1$ is the magnetic dimension $X_h$, and $X_{\rm 2s}$ is the thermal dimension $X_t = d-1/\nu$. 
    Exponents $X_1$ and  $X_{\rm 2s}$ are from Ref.~\cite{Xu2014} for 3D,
    and  the $X_1$ values are taken from Ref.~\cite{zhongjin} for 4D and 5D. 
    The 2D and 3D results, except $X_1$ and $X_{\rm 2s}$, are from Ref.~\cite{tan2019observation}.
    Note that a meaningful error bar is unavailable for $X_{\rm 3a}$ in 5D.
    }
    \label{tab_sum}
\end{table}

We study  the $N$-cluster correlation functions for critical percolation in 4D and 5D. 
We reformulate the transfer MC algorithm in Ref.~\cite{deng2005} by
utilizing a disjoint-set data structure, and carry out extensive Monte Carlo simulations
for cylindrical systems of size $L^{d-1}\times \infty$. 
The linear system size is up to $L=512$ in 4D and 128 in 5D. 
From finite-size scaling, we report the estimates 
of all the $N=2$ and 3 exponents 
and the universal logarithmic amplitude $\delta$. 
Table~\ref{tab_sum} summarizes the $d$-dependence of these  exponents and the amplitude $\delta$.

For $N=1$, the only correlation function $P_1=\bbp(\Pone)$ describes the probability that two distant sites 
are in the same percolation cluster,  and the exponent $X_1$ is  the magnetic dimension $X_h=d-y_{\rm h}$,
which has been extensively studied in Refs.~\cite{ballesteros1997,Gracey2015,zhongjin,Mertens}.
The symmetric exponent $X_{\rm 2s}$ is the so-called two-arm exponent. 
For percolation--i.e., the $Q \rightarrow 1$ limit of the $Q$-state Potts model, 
it is identical to the thermal scaling dimension $X_{\rm t}=d-y_{\rm t}$.
Thus, our estimate of $X_{\rm 2s}$ also gives a determination of $y_{\rm t}$ for 4D and 5D,
which significantly improves over the existing results.
To our knowledge, the other exponents for $N=2$ and 3 
and the universal amplitude $\delta$  have not been reported  for 4D and 5D. 

For spatial dimensions $d > 2$, exact results are unavailable for critical phenomena, 
mainly due to the finiteness of the conformal group. 
The approach from logarithmic conformal field theory (LCFT) focuses on nonlocal observables rather than local ones.
Combining the symmetry group ${\cal S}_N$ of $N$ clusters and the ${\cal S}_Q$ symmetry for the $Q$-state Potts model, 
the LCFT approach gives exact structural properties on nonlocal connectivities and defines an infinite family 
of $N$-cluster correlation functions that have symmetry-dependent critical exponents. 
Together with Ref.~\cite{tan2019observation}, our work confirms the validity of LCFT 
and produces one of the scarce pieces of knowledges for high-dimensional percolation with $d<6$~\cite{AraujoEtAl14}.

Above the upper dimensionality $d \geq d_{\rm u} = 6$, the critical behavior is governed by the Gaussian fixed point 
in the framework of renormalization group. This implies that, at criticality, 
the random path connecting a pair of distant sites along a percolation cluster is effectively a simple random walk in the continuum limit.
As a consequence, an $N$-cluster connectivity event, 
that there are $N$ distinct clusters propagating from a neighborhood ${\cal V}_i$ to another one far away ${\cal V}_j$,  
can be regarded as the event that there are $N$ simple random walks  between ${\cal V}_i$ and ${\cal V}_j$. 
Since the random walks are independent to each other, we speculate that for $d \geq 6$,
the symmetric exponent simply takes value $ X_{N{\rm s}}=2N$, 
supported by the known results $X_1=2$ and $X_{\rm 2s}=4$.
Further, we speculate that, in the continuum limit,
the cancellation of connectivity probabilities in correlators $P_{N,{\rm o}}$ would be so complete 
that the $N$-cluster exponents $X_{N{\rm o}}$, except $X_{N{\rm s}}$, are ill-defined, 
denoted as symbol ``/" in Table~\ref{tab_sum}.
To confirm/falsify these speculations, theoretical insights are desired since numerical study is extremely challenging.  

Given a spatial dimensionality $d<6$, we conjecture that the symmetric exponents are superadditive, i.e.,
$X_{N{\rm s}} > X_{N'{\rm s}} +X_{N''{\rm s}}$
for any set of positive integers with $N=N'+N''$.
This concavity property immediately yields another inequality $X_{N{\rm s}} > NX_1$.
The conjecture holds true for $N=1,2,3$ and $d=2,3,4,5$, as shown in Table~\ref{tab_sum}.
Moreover, the ratio $X_{\rm 2s}/X_1$ and $X_{\rm 3s}/X_1$ increase as $d$ decreases.
The underlying argument is as follows.
As $d$ is lowered from $d_{\rm u}=6$, critical percolation clusters become more and more 
compact~\cite{zhongjin, Xu2014a, huangwei2018, Wang13}.
Consider the process of simultaneously growing $N$ distinct clusters from a neighborhood of $N$ lattice sites, 
the clusters are very likely to merge together if they do not die out, due to their geometric compactness/fatness.
Thus, in the unlikely event that all the $N$ clusters survive up to distance $r$, they would separate from each other as $r$ increases. 
This leads to our expectation that $\bbp_N \leq \bbp_{N'} \cdot \bbp_{N''}$ for any positive integers $N=N'+N''$. 

Finally, we remark on the $d$-dependent behavior of the universal logarithmic amplitude $\delta$. 
Unlike the $N$-cluster exponents $X_{N{\rm o}}$ that are a monotonically increasing function of $d$, 
the $\delta$ value starts from $2\sqrt{3}/\pi \approx 1.103 $ in 2D, reaches a maximum $1.52(3)$ in 3D, 
and then decreases to $1.16(1)$ for 4D and $0.74(6)$ for 5D. 
The universal amplitude $\delta$ characterizes both the LCFT at $Q = 1$ and the limit of conformal field theories when $Q \rightarrow 1$,
and it is predicted that~\cite{Vasseur2012} 
\begin{equation}
\delta = 2 \times \lim_{Q \rightarrow 1} \frac{X_2-X_{\rm t}}{Q-1} \; ,
\end{equation}
where the two-arm exponent $X_2$ collides with the thermal exponent $X_{\rm t}$ in the $Q \rightarrow 1$ limit.
On the complete graph, it has been rigorously proved~\cite{Luczak2006} that as long as $Q <2$, the Fortuin-Kasteleyn cluster representation 
of the $Q$-state Potts model  belongs to the same universality class as the mean-field percolation. 
We speculate that this would hold for dimension $d>6$. 
Thus, one has $X_2 = X_{\rm t}$ and $\delta=0$ for  $d \geq 6$, 
in line with the decreasing tendency of $\delta$ as $d$ increases.
The value $\delta=0$ might indicate that for $d \geq 6$, the logarithmic correlation function $F(r)$ 
no longer has logarithmic $r$-dependence.

\acknowledgments 
We dedicate this work to Fred (Fa-Yueh) Wu who passed away on January 21, 2020. 
Known internationally for his contributions in statistical mechanics and solid state physics, Wu was a professor at Northeastern 
University for 39 years until his retirement in 2006 as Matthews Distinguished University Professor of Physics. 
His seminal review article on the Potts model~\cite{FYWu} has benefitted several generations of statistical physicists.
His broad interests in influence on his research community were illustrated by the special issue~\cite{jesper2012} that one of us (JLJ) 
co-edited for his 80-year birthday. In 2004,
Wu was a member of the doctoral dissertation committee of another of us (YD), and subsequently gave him a lot of encouragement 
throughout his academic career.

We are indebted to Romain Couvreur for valuable discussions.
YD acknowledges the support
by National Natural Science Foundation of China (Grant No.~11625522) and the Ministry of Science and Technology of
China (Grant No.~2016YFA0301604).
JLJ acknowledges support of the European Research Council through the Advanced Grant NuQFT.
Simulations were carried out at the Supercomputing Center of the University of Science
and Technology of China.

\input{mybib}
\end{document}

%% file: mycmd.tex
\def\bbp{{\mathbb P}}
\def\bfr{{\mathbf r}}

\newcommand{\np}{\;\;}
\newcommand*{\rom}[1]{\expandafter \romannumeral #1}

\newcommand{\Pone}{
\begin{tikzpicture}[baseline=1pt]
 \draw [fill] (0,0) circle(0.2ex);
 \draw [fill] (0,0.25) circle(0.2ex);
 \draw [thick] (0,0)--(0,0.25);
\end{tikzpicture}
}

\newcommand{\PPzero}{
\begin{tikzpicture}[baseline=1pt]
 \draw [fill] (0,0) circle(0.2ex);
 \draw [fill] (0.1,0) circle(0.2ex);
 \draw [fill] (0,0.25) circle(0.2ex);
 \draw [fill] (0.1,0.25) circle(0.2ex);
\end{tikzpicture}
}

\newcommand{\PPca}{
\begin{tikzpicture}[baseline=1pt]
 \draw [fill] (0,0) circle(0.2ex);
 \draw [fill] (0.1,0) circle(0.2ex);
 \draw [fill] (0,0.25) circle(0.2ex);
 \draw [fill] (0.1,0.25) circle(0.2ex);
 \draw [thick] (0,0)--(0,0.25);
\end{tikzpicture}
}

\newcommand{\PPcb}{
\begin{tikzpicture}[baseline=1pt]
 \draw [fill] (0,0) circle(0.2ex);
 \draw [fill] (0.1,0) circle(0.2ex);
 \draw [fill] (0,0.25) circle(0.2ex);
 \draw [fill] (0.1,0.25) circle(0.2ex);
 \draw [thick] (0.1,0)--(0.1,0.25);
\end{tikzpicture}
}

\newcommand{\PPcc}{
\begin{tikzpicture}[baseline=1pt]
 \draw [fill] (0,0) circle(0.2ex);
 \draw [fill] (0.1,0) circle(0.2ex);
 \draw [fill] (0,0.25) circle(0.2ex);
 \draw [fill] (0.1,0.25) circle(0.2ex);
 \draw [thick] (0,0)--(0.1,0.25);
\end{tikzpicture}
}

\newcommand{\PPcd}{
\begin{tikzpicture}[baseline=1pt]
 \draw [fill] (0,0) circle(0.2ex);
 \draw [fill] (0.1,0) circle(0.2ex);
 \draw [fill] (0,0.25) circle(0.2ex);
 \draw [fill] (0.1,0.25) circle(0.2ex);
 \draw [thick] (0,0.25)--(0.1,0);
\end{tikzpicture}
}

\newcommand{\PPa}{
\begin{tikzpicture}[baseline=1pt]
 \draw [fill] (0,0) circle(0.20ex);
 \draw [fill] (0.10,0) circle(0.20ex);
 \draw [fill] (0,0.25) circle(0.20ex);
 \draw [fill] (0.10,0.25) circle(0.20ex);
 \draw [thick] (0,0)--(0,0.25);
 \draw [thick] (0.10,0)--(0.10,0.25);
\end{tikzpicture}
}

\newcommand{\PPb}{
\begin{tikzpicture}[baseline=1pt]
 \draw [fill] (0,0) circle(0.2ex);
 \draw [fill] (0.1,0) circle(0.2ex);
 \draw [fill] (0,0.25) circle(0.2ex);
 \draw [fill] (0.1,0.25) circle(0.2ex);
 \draw[thick] (0,0)--(0.1,0.25);
 \draw[thick] (0.1,0)--(0,0.25);
\end{tikzpicture}
}

\newcommand{\PPPa}{
\begin{tikzpicture}[baseline=1pt]
 \draw [fill] (0,0) circle(0.2ex);
 \draw [fill] (0.1,0) circle(0.2ex);
 \draw [fill] (0.2,0) circle(0.2ex);
 \draw [fill] (0,0.25) circle(0.2ex);
 \draw [fill] (0.1,0.25) circle(0.2ex);
 \draw [fill] (0.2,0.25) circle(0.2ex);
  \draw [thick] (0,0)--(0,0.25);
  \draw [thick] (0.1,0)--(0.1,0.25);
  \draw [thick] (0.2,0)--(0.2,0.25);
\end{tikzpicture}
}

\newcommand{\PPPb}{
\begin{tikzpicture}[baseline=1pt]
 \draw [fill] (0,0) circle(0.2ex);
 \draw [fill] (0.1,0) circle(0.2ex);
 \draw [fill] (0.2,0) circle(0.2ex);
 \draw [fill] (0,0.25) circle(0.2ex);
 \draw [fill] (0.1,0.25) circle(0.2ex);
 \draw [fill] (0.2,0.25) circle(0.2ex);
  \draw [thick] (0,0)--(0,0.25);
  \draw [thick] (0.1,0)--(0.2,0.25);
  \draw [thick] (0.2,0)--(0.1,0.25);
\end{tikzpicture}
}

\newcommand{\PPPc}{
\begin{tikzpicture}[baseline=1pt]
 \draw [fill] (0,0) circle(0.2ex);
 \draw [fill] (0.1,0) circle(0.2ex);
 \draw [fill] (0.2,0) circle(0.2ex);
 \draw [fill] (0,0.25) circle(0.2ex);
 \draw [fill] (0.1,0.25) circle(0.2ex);
 \draw [fill] (0.2,0.25) circle(0.2ex);
  \draw [thick] (0,0)--(0.1,0.25);
  \draw [thick] (0.1,0)--(0.0,0.25);
  \draw [thick] (0.2,0)--(0.2,0.25);
\end{tikzpicture}
}

\newcommand{\PPPd}{
\begin{tikzpicture}[baseline=1pt]
 \draw [fill] (0,0) circle(0.2ex);
 \draw [fill] (0.1,0) circle(0.2ex);
 \draw [fill] (0.2,0) circle(0.2ex);
 \draw [fill] (0,0.25) circle(0.2ex);
 \draw [fill] (0.1,0.25) circle(0.2ex);
 \draw [fill] (0.2,0.25) circle(0.2ex);
  \draw [thick] (0,0)--(0.1,0.25);
  \draw [thick] (0.1,0)--(0.2,0.25);
  \draw [thick] (0.2,0)--(0.0,0.25);
\end{tikzpicture}
}

\newcommand{\PPPe}{
\begin{tikzpicture}[baseline=1pt]
 \draw [fill] (0,0) circle(0.2ex);
 \draw [fill] (0.1,0) circle(0.2ex);
 \draw [fill] (0.2,0) circle(0.2ex);
 \draw [fill] (0,0.25) circle(0.2ex);
 \draw [fill] (0.1,0.25) circle(0.2ex);
 \draw [fill] (0.2,0.25) circle(0.2ex);
  \draw [thick] (0,0)--(0.2,0.25);
  \draw [thick] (0.1,0)--(0.0,0.25);
  \draw [thick] (0.2,0)--(0.1,0.25);
\end{tikzpicture}
}

\newcommand{\PPPf}{
\begin{tikzpicture}[baseline=1pt]
 \draw [fill] (0,0) circle(0.2ex);
 \draw [fill] (0.1,0) circle(0.2ex);
 \draw [fill] (0.2,0) circle(0.2ex);
 \draw [fill] (0,0.25) circle(0.2ex);
 \draw [fill] (0.1,0.25) circle(0.2ex);
 \draw [fill] (0.2,0.25) circle(0.2ex);
  \draw [thick] (0,0)--(0.2,0.25);
  \draw [thick] (0.1,0)--(0.1,0.25);
  \draw [thick] (0.2,0)--(0.0,0.25);
\end{tikzpicture}
}

\usepackage{amssymb}
\usepackage{tikz}
\usetikzlibrary{calc}
\usetikzlibrary{chains}
\usetikzlibrary{scopes}
\usetikzlibrary{shapes.geometric}
\usetikzlibrary{trees}
\usetikzlibrary{topaths}
\usetikzlibrary{positioning}
\usetikzlibrary{patterns,decorations.pathreplacing}

\newcommand{\bP}{\mathbb P}

